\documentclass[a4paper,11pt]{article}
\pdfoutput=1 
\usepackage{amsmath}
\usepackage{amsfonts}
\usepackage{amssymb}
\usepackage{graphicx}%
\usepackage{amsthm}
\usepackage{mathrsfs}
\newcommand{\A}{\mathscr{A}}
\newcommand{\B}{\mathscr{B}}
\newcommand{\C}{\mathscr{C}}
\newcommand{\M}{\mathscr{M}}
\newcommand{\D}{\mathscr{D}}
\newcommand{\G}{\mathscr{G}}

\usepackage{jheppub} 
\usepackage[T1]{fontenc} 

\title{\boldmath T-duality/plurality of BTZ black hole metric coupled to two fermionic fields}

\author[1]{Ali Eghbali\note{Corresponding author.},}
\author[]{Meysam Hosseinpour-Sadid,}
\author[]{Adel Rezaei-Aghdam}

\affiliation[]{Department of Physics, Faculty of Basic Sciences,\\Azarbaijan Shahid Madani University, 53714-161, Tabriz, Iran}

\emailAdd{eghbali978@gmail.com}
\emailAdd{meysam.hs.az@gmail.com}
\emailAdd{rezaei-a@azaruniv.ac.ir}

\keywords{Sigma Models, String Duality, Conformal and W Symmetry}

\abstract{We ask the question of classical super (non-)Abelian T-duality for BTZ black hole metric coupling to two fermionic fields.
Our approach is based on super Poisson-Lie (PL) T-duality
in the presence of spectator fields.
In order to study the Abelian T-duality of the metric we dualize over the Abelian Lie supergroups of the types $(1|2)$ and $(2|2)$, in such a way that
it is shown that both original and dual backgrounds of the models are conformally invariant up to one-loop order in the presence of field strength.
Then, we study the non-Abelian T-duality of the BTZ vacuum metric coupling to two fermionic fields.
The dualizing is performed on some non-Abelian Lie supergroups of the type $(2|2)$,
in such a way that we are dealing with semi-Abelian superdoubles which are non-isomorphic as Lie superalgebras in each of the models.
In the non-Abelian T-duality case, it is interesting to mention that the models can be conformally invariant up to one-loop order in
both cases of the absence and presence of field strength.
In addition, starting from the decomposition of semi-Abelian Drinfeld superdoubles generated by some of the ${\C}^3 \oplus {\A}_{1,1}$ Lie superbialgebras
we study the super PL T-plurality of the BTZ vacuum metric coupled to two fermionic fields.
However, our findings are interesting in themselves, but at a constructive level, can prompt many new insights into supergravity and manifestly have interesting mathematical relationships
with double field theory.\\\\
{\scriptsize {\it The paper is dedicated to the fond memory of professor Farhad Darabi \\[-2mm]who passed away on 13 Sept 2022.}}
}

\begin{document}
\maketitle
\flushbottom
\section{Introduction}
\label{Sec.I}
The BTZ black hole, named after Banados, Teitelboim, and Zanelli (BTZ), is a black hole solution for $2+1$-dimensional
topological gravity with a negative cosmological constant, mass, angular momentum
and charge \cite{Banados1}.
The BTZ black hole is asymptotically anti-de Sitter rather than asymptotically flat, and has no curvature singularity
at the origin, because \cite{Banados2} it is a discrete quotient of $SL(2, \mathbb{R})$.
The line element for the  black hole solutions is given by
\begin{eqnarray}\label{int.1}
ds^2  = ({M} - {  r^2\over l^2}) d t^2 -J dt d\varphi  + r^2 d \varphi^2 + \big({  r^2\over l^2} - {M}  +
{J^2\over 4r^2}\big)^{-1} d  r^2,
\end{eqnarray}
where the radius $l$ is related to the cosmological constant by $l = (-\Lambda)^{-1/2}.$ Here,
the coordinate $\varphi$ is periodic with period $2 \pi$.
The constants of motion ${M}$ and $J$ are the mass and angular momentum of the BTZ black hole, respectively.
They are appeared due to the time translation symmetry and rotational symmetry of the metric,
corresponding to the killing vectors ${\partial / \partial t}$ and ${\partial/\partial \varphi}$, respectively.
The line element \eqref{int.1} describes a black hole solution with outer and inner horizons
at $r =r_+$ and $r= r_-$, respectively,
\begin{eqnarray}\label{b.3}
r_{\pm }= l \bigl ({{M}\over 2}\bigr )^{1/2}
\biggl \{1\pm \biggl (1- {{J^2} \over {{M^2} l^2}} \biggr)^{1/2}\biggr \}^{1/2}.
\end{eqnarray}
where the mass and angular momentum are related to $r_{\pm }$ by
\begin{eqnarray}\nonumber
{M} = \frac{r_{+}^2 + r_-^2}{l^2},~~~~~~~~~~~J = \frac{2 r_{+}  r_-}{l}.
\end{eqnarray}
The region $r_+  < r < {M}^{1/2} l$ defines an ergosphere, in which the
asymptotic timelike Killing field $\partial/ {\partial t}$ becomes spacelike.
The solutions with  $- 1 < {M} < 0, \;J=0$  describe  point
particle sources with naked conical singularities at
$r=0  $. The  metric with ${M}= -{1 },\; J=0 $ may be recognized as that of ordinary anti-de Sitter space; it
is separated by a mass gap from the ${M}= 0,\; J=0 $ ``massless black hole'',
whose geometry is discussed in Ref. \cite{Banados2}.
The vacuum state which is regarded as empty space,  is obtained by letting the horizon size go to zero. This
amounts to letting ${M} \rightarrow 0$, which requires $J \rightarrow 0$. We have to
notice that the metric for the ${ M} = J = 0$ black hole is not the same as $AdS_3$ metric which has negative mass ${M}=-1$.
Locally they are equivalent since there is locally only one constant curvature metric in three dimensions.
However they are inequivalent globally.
Notice that the singularity at the center of the BTZ black hole is
called a conical singularity. However, a conical singularity is not
similar to a canonical singularity because it does not cause the
spacetime curvature to diverge \cite{Ning}.\\
$~~~~$The BTZ black hole also arises in string theory as a near-horizon region of a non-extremal system of
1- and 5-branes. In Ref. \cite{Skenderis} by using a presentation of the D-brane counting for extremal black
holes, it has been shown that four- and five-dimensional non-extremal black holes can be mapped to the
BTZ black hole (times a compact manifold) by means of dualities.
A slight modification of this black hole solution
yields an exact solution to string theory \cite{Horowitz2}.
There, the BTZ black hole solutions have been considered in the context of the
low energy approximation, then as the exact conformal field theory.
In order to obtain an exact solution to string theory, one must modify the BTZ black hole solutions
by adding an antisymmetric tensor field $H_{_{MNP}}$ (field strength) proportional to the volume form $\varepsilon_{_{MNP}}$.
It has been shown that any solution to three-dimensional
general relativity with negative cosmological constant is a solution to low energy string theory with
$H_{_{MNP}} = 2\epsilon_{_{MNP}}/l$, $\Lambda = -1/l^2$ and a dilaton field $\Phi =0$ \cite{Horowitz2}.
In particular, it was observed in \cite{Horowitz2} that, the two parameter
family of black holes \eqref{int.1} along with
\begin{eqnarray}
B_{\varphi t} = \frac{r^2}{l},~~~~~~~~~~~\Phi =0,\label{b.6}
\end{eqnarray}
satisfy the low energy string effective action equations. Then, it was obtained \cite{Horowitz2}
the Abelian dual of this solution by Buscher's duality transformations \cite{Buscher1}.
Indeed,  T-duality symmetry is one of the most interesting properties of string theory
that connects seemingly different backgrounds in which
the strings can propagate.
The duality is Abelian if it is constructed on an Abelian isometry group.
We refer to Abelian T-duality for stressing the presence of global Abelian isometries in the target spaces of both the paired $\sigma$-models, while
non-Abelian T-duality \cite{de.laOssa} refers to the existence of a non-Abelian isometry on the target space of one of the two
$\sigma$-models.
Now that we are discussing the non-Abelian T-duality, it is necessary to explain that the non-Abelian target space
dual of BTZ vacuum solution was obtained in \cite{EMR13} by making use of the PL T-duality approach in the presence
of spectator fields. There, it has been shown that the BTZ vacuum solution
with no horizon and no curvature singularity is related to
a solution with a single horizon and a curvature singularity.
The PL T-duality was proposed by Klimcik and Severa \cite{{Klim1},{Klim2}} as a
generalization of Abelian \cite{Buscher1} and non-Abelian dualities
\cite{de.laOssa}. In this type of duality, the symmetry does not need
to be realized as an isometry of the initial background; moreover, $\sigma$-models
are formulated on a Drinfeld double group $D$ \cite{Drinfeld}
whose Lie algebra $\D$ can be decomposed into a pair of maximally isotropic
sub-algebras with respect to a non-degenerate invariant bilinear form on $\D$,
so that the sub-algebras are duals of each other in the usual sense.
In this case it is said that the $\sigma$-models are indeed dual in the sense of the PL T-duality.\\
$~~~~$Generalization of T-duality to Lie supergroups and  also supermanifolds have been explored in the context of
PL T-duality in \cite{{ER2},{ER5}}.
When we deal with a semi-Abelian Drinfeld superdouble, that is, one of the sub-supergroups of Drinfeld superdouble is considered to be Abelian,
the super PL T-duality reduces to the super non-Abelian T-duality.
Recently, it has been considered  the non-Abelian T-duality of $\sigma$-models on group
manifolds, symmetric, and semi-symmetric spaces with the emphasis on the T-dualization of
$\sigma$-models whose target spaces are supermanifolds \cite{Bielli1} (see also \cite{Bielli2}). There, it has been performed super non-Abelian T-dualization
of the principal chiral model based on the $OSP(1|2)$ Lie supergroup.\\
$~~~~$The first goal of this paper is to find (Abelian)non-Abelian target space duals of the BTZ metric when is coupled to two fermionic fields.
In this regards, the target spaces of T-dual $\sigma$-models are considered to be
the supermanifolds ${\M} \approx {O} \times { G}$ and ${\tilde \M} \approx {O} \times { \tilde G}$
where ${G}$ and $ { \tilde G}$ are
(Abelian)non-Abelian Lie supergroups of the type $(2|2)$ acting freely on ${\M}$ and  ${\tilde \M} $, respectively,
while ${O}$ is the orbit of $ G$ in ${\M}$.
In the Abelian T-duality case, we also dualize over the Abelian Lie supergroup of the type $(1|2)$.
The indecomposable four-dimensional Lie superalgebras of the type $(2|2)$ was first classified by Backhouse in \cite{B}.
Unfortunately, there are some omissions and redundancies in that paper, probably because
the author in some cases has not taken into account all isomorphisms to reduce the list of superalgebras.
We have made some modifications to the Backhouse's classification; the detail of the calculations are given in Appendix A.
In addition, one can find a classification of decomposable Lie superalgebras of the type $(2|2)$ in \cite{ER6}.
Based on the (anti-)commutation relations between bases of boson-boson (B-B), boson-fermion (B-F) and fermion-fermion (F-F)
we have classified all decomposable and indecomposable Lie superalgebras of the type $(2|2)$ into six disjoint families in Appendix A.
Finally, we find super PL T-plurals of the BTZ vacuum metric coupled to two fermionic fields with
respect to the ${C}^3 \oplus {A}_{1,1}$ Lie supergroup.
The general procedure that we shall apply is a straightforward generalization of
the well-known prescription of purely bosonic PL T-plurality \cite{vonUnge} to Lie supergroup case, and
we shall refer to it as super PL T-plurality in this work.
Notice that the generalization of PL T-plurality formulation to Lie supergroup case
has been recently explored in \cite{Eghbali3}.\\
$~~~~$The outline of this paper is as follows: in section \ref{Sec.II} we introduce our
notations and recall the origin of the super PL T-duality transformations in the presence of spectator fields at the level of the $\sigma$-model.
Abelian and non-Abelian target space duals of the BTZ metric coupled to two fermionic fields are investigated in sections \ref{Sec.III}
and \ref{Sec.IV}, respectively; the results of non-Abelian T-duality of the models are summarized in Table 1 at the end of section \ref{Sec.IV}.
In section \ref{Sec.V}, we study super PL T-plurality of the BTZ vacuum metric coupled to two fermionic fields,
in such a way that we start from the decompositions of semi-Abelian Drinfeld superdoubles generated by the ${\C}^3 \oplus {\A}_{1,1}$ Lie superbialgebras.
Finally, in section \ref{Sec.VI}, we present our conclusions and sketch possible developments of this work.
The classification of decomposable and indecomposable Lie superalgebras of the type $(2|2)$ is left to Appendix A.
Appendix B is also devoted to solutions of the super Jacobi and mixed super Jacobi
identities for the ${\C}^3 \oplus {\A}_{1,1}$ Lie superalgebra.

\section{A brief review of super PL T-duality with spectators}
\label{Sec.II}

Before proceeding to review the super PL T-duality on supermanifolds,
let us recall the properties of $\mathbb{Z}_2$-graded vector
space and also some definitions related to Lie superalgebras \cite{N.A}.
A super vector space $\mathbb{V}$ is a ${\mathbb{Z}}_{2}$-graded vector space, i.e., a vector space over a field
$\mathbb{K}$ with a given decomposition of sub-spaces of grade 0 and grade 1, $\mathbb{V}= \mathbb{V}_{_0} \oplus \mathbb{V}_{_1}$.
The parity of a nonzero homogeneous element, denoted by $|x|$, is $0$ (even) or
$1$ (odd)\footnote{The even elements are sometimes called bosonic, and the odd elements fermionic. From now on,
we use $B$ and $F$ instead of $0$ and $1$, respectively.} according to whether it is in $\mathbb{V}_{0}$ or $\mathbb{V}_{1}$, namely,
$|x|=0$ for any $x \in \mathbb{V}_{_0}$, while for any $x \in \mathbb{V}_{_1}$ we have $|x|=1$.
A {\it Lie superalgebra} ${\G}$ is a $\mathbb{Z}_2$-graded vector space, thus admitting the decomposition
${\G} ={\G}_{{_B}} \oplus {\G}_{_F}$, equipped with a bilinear
superbracket structure $[. , .]: {\G} \otimes {\G} \rightarrow {\G}$ satisfying the requirements of anti-supersymmetry
and super Jacobi identity.
If ${\G}$ is finite-dimensional and the dimensions of ${\G}_{_B}$ and $ {\G}_{_F}$
are $m=\#B$ and  $n=\#F$, respectively, then ${\G}$ is said to have dimension $(m|n)$.
We shall identify grading indices by the same indices in the power of $(-1)$, i.e., we use $(-1)^x$ instead of  $(-1)^{|x|}$, where
$(-1)^x$ equals 1 or -1 if the Lie sub-superalgebra element is even or odd, respectively\footnote{
Note that this notation was first used by DeWitt in \cite{D}. Throughout this paper we work with DeWitt's notation.}.

Since our main discussion will be related to super PL T-duality, it is worth mentioning that
super PL T-duality is based on the concepts of Drinfeld superdouble \cite{{N.A},{ER1},{ER4}}, so it is necessary to define Drinfeld superdouble supergroup $D$.
Following \cite{Drinfeld}, a Drinfeld superdouble is simply a Lie supergroup $D$ whose Lie superalgebra $\D$
admits a decomposition $\D =\G \oplus {\tilde \G}$ into a pair of sub-superalgebras maximally isotropic
with respect to a supersymmetric ad-invariant non-degenerate bilinear form $<.~,~.>$.
The dimension of sub-superalgebras have to be equal. We furthermore consider $G$ and $\tilde G$ as a pair of maximally isotropic sub-supergroups corresponding
to the sub-superalgebras $\G$ and $\tilde \G$,
and choose a basis in each of the sub-superalgebras as
$T_{{_a}} \in \G$  and ${\tilde T}^{{^a}} \in {\tilde \G}, a = 1, ..., dim~G$, such that
\begin{eqnarray}\label{3.1}
<T_{{_a}} ,  T_{{_b}}> = <{\tilde T}^{{^a}} ,  {\tilde T}^{{^b}}> =0,~~~~~~~{\delta}_{_{a}}^{{~b}}=<T_{{_a}} ,
 {\tilde T}^{{^b}}>  = (-1)^{ab}~<{\tilde T}^{{^b}} , T_{{_a}} >.~~
\end{eqnarray}
The basis of the two sub-superalgebras satisfy the commutation relations
\begin{eqnarray}\label{3.2}
[T_a , T_b] = {f^{c}}_{ab} ~T_c,~~~~~[{\tilde T}^{a} , {\tilde T}^{b}] ={{\tilde f}^{ab}}_{~~c} ~{\tilde T}^{c},~~~~
[T_a , {\tilde T}^{b}] = (-1)^b {{\tilde f}^{bc}}_{~~a} {T}_c + (-1)^a {f^{b}}_{ca} ~{\tilde T}^{c},~~~
\end{eqnarray}
where ${f^{c}}_{ab}$ and $\tilde f^{ab}_{~~c}$ are structure constants of $\G$ and $\tilde \G$, respectively.
Noted that the Lie superalgebra structure defined by relation \eqref{3.2} is called {\it Drinfeld superdouble} $\D$.
The super Jacobi identity on $\D$ relates the structure constants of the two Lie
superalgebras as \cite{ER1}
\begin{eqnarray}\label{3.3}
{f^e}_{bc}{\tilde{f}^{ad}}_{\; \; \; \; e}=
{f^a}_{ec}{\tilde{f}^{ed}}_{\; \; \; \; \; b} +
{f^d}_{be}{\tilde{f}^{ae}}_{\; \; \; \; \; c}+ (-1)^{bd}
{f^a}_{be}{\tilde{f}^{ed}}_{\; \; \; \; \; c}+ (-1)^{ac}
{f^d}_{ec}{\tilde{f}^{ae}}_{\; \; \; \; \; b}.
\end{eqnarray}

To continue, let us now consider a two-dimensional non-linear $\sigma$-model for the $d$ field variables
$X^{^{M}} = (y^i, x^{\mu})$, where $x^\mu,~ \mu = 1, \cdots, dim\hspace{0.4mm}G$ stand for the coordinates of Lie supergroup $G$ acting freely from right on
the target supermanifold $\M \approx O \times G$, and $y^i,~i = 1, \cdots , d-dim\hspace{0.4mm}G$ are
the coordinates labeling the orbit $O$ of $ G$ in  ${\M}$. Here we work in
the standard light-cone variables on the worldsheet ${\Sigma}$,
$\sigma^{\pm} =(\tau \pm \sigma)/{2}$ together with $\partial_{\pm}=\partial_{\tau} \pm \partial_{\sigma}$.
It should be remarked that the coordinates $y^i$ do not participate in the PL T-duality transformations
and are therefore called spectators \cite{Sfetsos1}.
The corresponding action has the form\footnote{As we will see, the spectators are considered to be bosonic (even) coordinates, therefore
$(-1)^i$ equals 1. }\cite{ER5} (see also \cite{{Klim1},{Klim2},{Sfetsos1},{Tyurin}})
\begin{eqnarray}\label{3.4}
S &=& \frac{1}{2} \int_{_{\Sigma}} d\sigma^{+}  d\sigma^{-} \Big[(-1)^a  R_{+}^a~{{E}_{_{ab}}}(g, y^i)~
 R_{-}^b + (-1)^a  R_{+}^a~ \phi^{{(1)}}_{a j}(g, y^i) \partial_{-} y^{j}\nonumber\\
&&~~~~~~~~~~~~~~~~~~~~~~~~~+ (-1)^i \partial_{+} y^{i} \phi^{{(2)}}_{i b}(g, y^i)  R_{-}^b + (-1)^i~ \partial_{+} y^{i} \phi_{_{ij}}(g, y^i) \partial_{-} y^{j} \Big].~~
\end{eqnarray}
where $R_{\pm}^a$ are the components of the right-invariant Maurer-Cartan super one-forms which are constructed by means of
an element $g$ of ${G}$ as follows:
\begin{eqnarray}\label{3.5}
R_{\pm} =(-1)^a  R_{\pm}^a~ T_a = (-1)^a (\partial_{\pm} g g^{-1})^a ~ T_a,
\end{eqnarray}
in which ${T}_a$, $a=1, \cdots, dim~G$ are the bases of the Lie superalgebra ${\G}$ of ${G}$.
The couplings ${{E}_{_{ab}}}, \phi^{{(1)}}_{a j}, \phi^{{(2)}}_{i b} $ and $\phi_{_{ij}}$ may depend on all variables $x^\mu$ and $y^i$.

Similarly we introduce another $\sigma$-model for the $d$ field variables ${\tilde X}^{^{M}} =({\tilde x}^{\mu} , y^i)$, where
${\tilde x}^{\mu}$'s, $\mu = 1, \cdots, dim\hspace{0.4mm}G$ parameterize an element ${\tilde g}$
of the dual Lie supergroup ${\tilde G}$, whose dimension is, however,
equal to that of $G$, and the rest of the variables are the same $y^i$'s used in \eqref{3.4}. Accordingly, we
introduce a different set of bases ${{\tilde T}^a }$ of the Lie algebra ${\tilde \G}$, with $a = 1, 2, \cdots , dim\hspace{0.4mm} G$.
We furthermore consider the components of the right-invariant Maurer-Cartan super one-forms on
${\tilde G}$ in the following form
\begin{eqnarray}\label{3.6}
(\partial_{\pm} \tilde g \tilde g^{-1})_a={\tilde R}_{{\pm}_a}=(-1)^\mu \partial_{\pm} {\tilde x}^{\mu} {\tilde R}_{\mu a}.
\end{eqnarray}
In this case, the corresponding action has the form
\begin{eqnarray}\label{3.7}
\tilde S &=& \frac{1}{2} \int_{_{\Sigma}} d\sigma^{+}  d\sigma^{-}\Big[(-1)^b {\tilde R}_{+_{a}} {{{\tilde E}}^{{ab}}}({\tilde g}, {y}^{i})~
{\tilde R}_{-_{b}}+{\tilde R}_{+_{a}} {\tilde \phi}^{\hspace{0mm}{(1)^{ a}}}_{~~~j}({\tilde g}, {y}^{i}) ~\partial_{-} y^{j}\nonumber\\
&&~~~~~~~~~~~~~~~~~ +(-1)^i \partial_{+} y^{i} {\tilde \phi}^{\hspace{0mm}{(2)^{ b}}}_{i}({\tilde g}, {y}^{i})  ~{\tilde R}_{-_{b}}
+ (-1)^i \partial_{+} y^{i} {\tilde \phi}_{_{ij}}({\tilde g}, {y}^{i}) \partial_{-} y^{j}\Big].
\end{eqnarray}
Notice that here one does not require any isometry associated with the Lie supergroups $G$ and $\tilde G$.
The $\sigma$-models \eqref{3.4} and \eqref{3.7} will be dual to each other in the sense of super PL
T-duality \cite{{ER2},{ER5}}  if the associated Lie superalgebras $\G$ and ${\tilde \G}$
form a Drinfeld superdouble which can be decomposed.

There remains to relate the couplings ${{E}_{_{ab}}}, \phi^{{(1)}}_{a j}, \phi^{{(2)}}_{i b} $
and $\phi_{_{ij}}$ in  \eqref{3.4} to ${{{\tilde E}}^{{ab}}}, {\tilde \phi}^{\hspace{0mm}{(1)^{ a}}}_{~~~j},
{\tilde \phi}^{\hspace{0mm}{(2)^{ b}}}_{i}$ and ${\tilde \phi}_{_{ij}}$ in \eqref{3.6}.
It has been shown that \cite{ER5} the various couplings in the
$\sigma$-model action \eqref{3.4} are restricted to be
\begin{eqnarray}\label{3.8}
{E_{_{ab}}} &=& {\big(E^{{-1}}_{0}+ \Pi\big)_{{ab}}}^{\hspace{-3mm}-1},~~~~~~~~~~~~~~~
\phi^{{(1)}}_{a i} = (-1)^c~{E_{_{ab}}}~({E^{{-1}}_{0}})^{bc}~{F_{~ci}}^{^{\hspace{-2mm}(1)}},\nonumber\\
\phi^{{(2)}}_{i b}&=& (-1)^c~{F_{~ia}}^{^{\hspace{-2.5mm}(2)}}~ ({E^{{-1}}_{0}})^{ac}~{E_{_{cb}}},~~
\phi_{_{ij}} = F_{_{ij}} - (-1)^{e+c}~{F_{~ia}}^{^{\hspace{-2.5mm}(2)}}~{\Pi}^{ac} {E_{_{cd}}} ~({E^{{-1}}_{0}})^{de}~{F_{~ej}}^{^{\hspace{-2mm}(1)}},~~~~~
\end{eqnarray}
where the new couplings $E_{0}, F^{^{(1)}}, F^{^{(2)}}$ and $F$  maybe functions of
the spectator variables $y^{i}$ only. In equation \eqref{3.8},
the Poisson superbracket $\Pi(g)$ is $\Pi^{^{ab}}(g) =(-1)^c  ~b^{^{ac}}(g)~ (a^{-1})_{_{c}}^{^{~b}}(g)$ so that sub-matrices
$a_{_{a}}^{^{~b}}(g)$ and $b^{^{ab}}(g)$ are defined in the following way
\begin{eqnarray}\label{3.10}
g^{-1} T_{{_a}}~ g &=& (-1)^b ~a_{_{a}}^{^{~b}}(g) ~ T_{{_b}},\nonumber\\
g^{-1} {\tilde T}^{{^a}} g &=&(-1)^c~ b^{^{ac}}(g)~ T_{{_c}}+{(a^{-st})^{^a}}_{_{b}}(g)~{\tilde T}^{{^b}}.
\end{eqnarray}
Finally, the relation of the dual action couplings to those of the original one is given by \cite{ER5}
\begin{eqnarray}
{{\tilde E}^{^{ab}}} &=& {\big(E_{0}+ {\tilde \Pi}\big)^{\hspace{-1mm}-1}}^{ab},~~~~~~~~~~
{\tilde \phi}^{\hspace{0mm}{(1)^{ a}}}_{~~~j} = (-1)^b~ {{\tilde E}^{^{ab}}}~{F_{~bj}}^{^{\hspace{-2.5mm}(1)}},\nonumber\\
{\tilde \phi}^{\hspace{0mm}{(2)^{ b}}}_{i} &=& - {F_{~ia}}^{^{\hspace{-2.75mm}(2)}} ~{{\tilde E}^{^{ab}}},~~~~~~~~~~~~~~~
{\tilde \phi}_{_{ij}} = F_{_{ij}}- (-1)^{b}~ {F_{~ia}}^{^{\hspace{-2.75mm}(2)}} ~{{\tilde E}^{^{ab}}}~{F_{~bj}}^{^{\hspace{-2mm}(1)}}.\label{3.11}
\end{eqnarray}
Analogously, one can define ${\tilde \Pi} (\tilde g)$ and sub-matrices ${\tilde a} (\tilde g), {\tilde b} (\tilde g)$ by just replacing the untilded symbols
by tilded ones.

Equivalently the action \eqref{3.4} can be expressed as
\begin{eqnarray}\label{3.12}
S = \frac{1}{2} \int_{_{\Sigma}} d\sigma^{+}  d\sigma^{-} (-1)^M~\partial_+ X^M~ {\cal E}_{_{MN}}(X)~\partial_- X^N,
\end{eqnarray}
where ${\cal E}_{_{MN}}(X)$ as a second order tensor field on the supermanifold $\M$ is a composition of the supersymmetric metric ${G}_{_{MN}}(X)$
and the anti-supersymmetric torsion potential ${B}_{_{MN}}(X)$ ($B$-field)\footnote{Note that $G_{_{MN}}$
and $B_{_{MN}}$ are  the components of the supersymmetric metric $G$ and the
anti-supersymmetric tensor field $B$, respectively,
$$
G_{_{MN}}\;=\;(-1)^{^{MN}}\;G_{_{NM}},~~~~~~~~~~~B_{_{MN}}\;=\;-(-1)^{^{MN}}\;B_{_{NM}}.
$$
We will assume that the metric ${_{_{M}}G_{_{N}}}$ is
superinvertible and its superinverse is shown by
$G^{^{MN}}$.}.
In the absence of spectator fields, the $ {\cal E}_{_{MN}}(X)$ reduces to $ {\cal E}_{_{\mu\nu}}(x)$. The relationship
between $ {\cal E}_{_{\mu\nu}}(x)$ and ${{E}_{_{ab}}}(g)$ is given by the formula
\begin{eqnarray}\label{3.13}
{\cal E}_{_{\mu\nu}}(x) =  (-1)^a~ {R_{_\mu}}^{~a}~ {{E}_{_{ab}}}(g) ~{{(R^{st})}^b}_{_\nu},
\end{eqnarray}
where the superscript ``$st$'' stands for supertransposition of the matrix.
The condition of super PL symmetry of $\sigma$-models on the level of the Lagrangian is
given by the formula \cite{ER2}
\begin{eqnarray}\label{3.14}
{\cal L}_{_{V_{_a}}}{\cal E}_{_{\mu\nu}}\;=\;(-1)^{^{a+\mu a+\rho c+\sigma}}~{{\tilde f}^{bc}}_{~~a}~ {\cal E}_{_{\mu \rho}}~ {V_{_c}}^{^{~\rho}}~{V_{_b}}^{^{~\sigma}}~
{\cal E}_{_{\sigma \nu}},
\end{eqnarray}
where ${{\tilde f}^{bc}}_{~~a}$ are structure coefficients of the dual Lie
superalgebra $\G$ and ${V_{_a}}^{^{~\mu}}$ are the components of left-invariant supervector fields
on the Lie supergroup $G$.
The superalgebras $\G$ and ${\tilde \G}$ then define the  Drinfeld superdouble that enables to
construct tensor ${\cal E}_{_{\mu \nu}}$ satisfying \eqref{3.14}.

\section{Abelian T-duality of BTZ metric coupled to two fermionic fields}
\label{Sec.III}

\subsection{Abelian T-duality with Abelian Lie supergroup of the type $(2|2)$}

In this subsection we construct explicitly a pair of Abelian T-dual $\sigma$-models on $4+1$-dimensional
supermanifolds ${\M}$ and $\tilde {\M}$ as the target superspaces.
The original model is built on the supermanifold ${\M} \approx {O} \times { G}$, where ${G}=I_{_{(2|2)}}$ is a
four-dimensional Abelian Lie supergroup of the type $(2|2)$ acting freely on ${\M}$ while ${O}$ is the orbit of $ G$ in ${\M}$.
The target superspace of the dual model is the supermanifold $\tilde {\M} \approx {O} \times \tilde {G}$, where
${\tilde {G}}$ is identical to ${G}$ acting freely on $\tilde {\M}$.
In this case, both the Lie supergroups $G$ and $\tilde {G}$ are Abelian, so we are dealing with a Abelian Drinfeld superdoubles of the type $(4|4)$.
Accordingly, by using \eqref{3.10} we conclude that both the Poisson superbracket $\Pi^{ab}(g)$ and ${\tilde \Pi}_{ab} {(\tilde g)}$ are vanished.
In what follows we shall show that the original model
describes a string propagating in a space with the BTZ black hole metric coupled to two fermionic fields.

\subsubsection{The original model}
In order to construct the original $\sigma$-model on the ${\M} \approx {O} \times { G}$ we assume that the Abelian Lie supergroup $G$ is parameterized
by an element
\begin{eqnarray}\label{3.15}
g= e^{t T_{_1}} e^{\varphi T_{_2}} e^{\psi T_{_3}} e^{\chi T_{_4}},
\end{eqnarray}
where $(t, \varphi)$  and  $(\psi, \chi)$ are the bosonic and fermionic coordinates, respectively; moreover,
$\{ T_{_1},  T_{_2}\}$ and $\{T_{_3}, T_{_4}\}$ stand for the bosonic and fermionic basis of Lie superalgebra $\G$ of $G$, respectively.
Here, the coordinate of orbit ${O}$ is represented by one
spectator field $y^i =\{r\}$. So, the coordinates of ${\M}$ are represented by $X^{^M}=(r, x^a)=(r, t, \varphi; \psi, \chi)$. In this case,
the components of the right-invariant super one-forms are simply obtained to be
$R_\pm^a =\partial_{\pm} x^a$.
Now, we need to determine the coupling matrices of the model. Let
us choose the spectator-dependent matrices in the following form
\begin{eqnarray}
{E_{0}}_{_{ab}}&=&\left( \begin{tabular}{cccc}
                 $M-\frac{r^2}{l^2}$ & $-\frac{J}{2}-k(r)$ & 0 & 0 \\
                 $-\frac{J}{2}+k(r)$ & $r^2$  & 0& 0 \\
                 0 & 0& $h_1(r)$ & $F(r)$ \\
                 0 & 0 & -$F(r)$ & $h_2(r)$ \\
                 \end{tabular} \right),\nonumber\\
F_{i j}  &=& \big({  r^2\over l^2} - {M}  +
{J^2\over 4r^2}\big)^{-1},~~~~~~~~~~~~~F^{(1)}_{a i}  = F^{(2)}_{i a} = 0,\label{3.16}
\end{eqnarray}
where $k(r), h_1(r), h_2(r)$ and $F(r)$ are some arbitrary functions that depend on the spectator field $\{r\}$ only.
As mentioned above, since the $G$ is Abelian, the Poisson superbracket $\Pi^{ab}(g)$ on $G$ is zero; consequently,
the various couplings in the action \eqref{3.4} can be reduced to
${{E}} = E_{0}, \phi^{{(1)}} = F^{^{(1)}}, \phi^{{(2)}} = F^{^{(2)}}$ and
$\phi = F $. Thus, by making use of the \eqref{3.4} and also the above results, the action of original $\sigma$-model is worked out to be
\begin{eqnarray}\label{3.17}
S&=& \frac{1}{2} \int d\sigma^{+}  d\sigma^{-} \Big[( { M} -{ r^2\over l^2}) \partial_{+}t ~ \partial_{-}t - \big({J\over{2}} + k(r)\big) \partial_{+}t ~ \partial_{-}{ \phi} -\big ({J\over{2}} - k(r)\big) \partial_{+}{ \phi}  ~\partial_{-}t~~~~~~~~~~\nonumber\\
&& +~ {r^2}~ \partial_{+}{ \phi}~ \partial_{-}{ \phi} - h_1(r)  \partial_{+}{ \psi}~  \partial_{-}{ \psi} - F(r) ~ \partial_{+}{\psi}~\partial_{-}{\chi}  +   F(r) ~ \partial_{+}{\chi}~\partial_{-}{\psi}- h_2(r)  \partial_{+}{ \chi}~  \partial_{-}{ \chi} \nonumber\\
&&  +~ ({  r^2\over l^2} - {M} +
{J^2\over 4r^2})^{-1}~\partial_{+}r~\partial_{-}r   \Big].
\end{eqnarray}
By considering the definition of line element and $B$-field on a supermanifold as
\begin{eqnarray}
ds^2 &=&(-1)^{^{MN}}~G_{_{MN}} ~ dX^M~dX^N,\label{3.18}\\
B&=&\frac{1}{2} (-1)^{^{MN}}~ B_{_{MN}}~dX^M  \wedge  dX^N,\label{3.19}
\end{eqnarray}
and then by comparing the actions \eqref{3.17} and \eqref{3.12} one gets
\begin{eqnarray}
ds^2&=& \big({r^2\over l^2} - M  +
{J^2\over 4r^2}\big)^{-1} dr^2 + (M -{ r^2\over l^2}) dt^2 - J dt d\phi + r^2 d\phi^2 - 2 F(r) d\psi d\chi,\label{3.20}\\
B&=& -k(r)~ dt \land d\phi - {1\over2} h_1(r)~ d\psi \land d\psi  - {1\over2} h_2(r)~ d\chi \land d\chi.\label{3.21}
\end{eqnarray}
In the $\sigma$-model context, the ultraviolet finiteness of the quantum version of the model is guaranteed
by the conformal invariance of the model. To achieve this invariance at the one-loop
level we must add another term containing the so-called dilaton field to the
Lagrangian of action \eqref{3.12}\footnote{In the super Abelian T-duality case,
the formula of dual dilaton transformation is  given by
\begin{eqnarray}
{\tilde \Phi} =\phi^{^{(0)}} - \frac{1}{2} \log |s\hspace{-0.6mm}\det ({{_a}({E_0})_b})|,\label{3.21.1}
\end{eqnarray}
where ``$s\hspace{-0.6mm}\det$'' stands for superdeterminant of the matrix; moreover,
 $\phi^{^{(0)}}$ is the dilaton that makes the original $\sigma$-model conformal up to the one-loop order
and may depend on both supergroup and spectator coordinates.}. The dilaton field $\Phi$ can be understood as an additional function on $\M$
that defines the quantum non-linear $\sigma$-model and couples to scalar curvature of
the worldsheet. The conformal invariance of the
model is guaranteed by vanishing of the so-called beta-function. At the one-loop level
the equations for vanishing of the beta-function on a supermanifold read \cite{ER5}\footnote{\emph{Notation:} Suppose that $f$ be a differentiable function on ${\mathbf{R}}_c^m \times {\mathbf{R}}_a^n$ (${\mathbf{R}}_c^m$ are subset of all real numbers with dimension $m$ while ${\mathbf{R}}_a^n$ are subset of all
odd Grassmann variables with dimension $n$), then,  the relation between the left partial differentiation
and right one is given by
$$
{_{_{M}}{\overrightarrow \partial}} f := \frac{\overrightarrow{\partial}f}{{\partial} X^{^{M}}}  \;=\; (-1)^{M(|f|+1)}\;  \frac{f \overleftarrow{\partial}}{{\partial} X^{^{M}}},
$$
in which $|f|$ stands for the grading of $f$ \cite{D}. }
\begin{eqnarray}
&&{\cal R}_{_{MN}}+\frac{1}{4}H_{_{MQP}} {H^{^{PQ}}}_{_{N}}+2{\overrightarrow{\nabla}_{_M}}
{\overrightarrow{\nabla}_{_N}} \Phi ~=~0,\label{3.22}\\
&&(-1)^{^{P}} {\overrightarrow{\nabla}}^{P}\big(e^{^{-2 \Phi}} H_{_{PMN}}\big)
~=~0,\label{3.23}\\
&&4 \Lambda-{\cal R}-\frac{1}{12} H_{_{MNP}}H^{^{PNM}} +4{\overrightarrow{\nabla}_{_M}} {\Phi} \overrightarrow{\nabla}^{^M} {\Phi}
 - 4 {\overrightarrow{\nabla}_{_M}} \overrightarrow{\nabla}^{^M} {\Phi}  =0,\label{3.24}
\end{eqnarray}
where the covariant derivatives ${\overrightarrow{\nabla}_{_M}}$, Ricci tensor ${\cal R}_{_{MN}}$
and scalar curvature ${\cal R}$ \footnote{One may refer to Appendix A of Ref. \cite{Eghbali3} to find a few relevant details concerning properties of matrices and tensors on supervector space which feature in the main text, appear as supertranspose, superdeterminant and supertrace, as well as the
definitions of connection, Riemann tensor field, Ricci tensor and scalar curvature on supermanifolds (for more details, refer to DeWitt's book \cite{D}).} are
calculated from the metric $~G_{_{MN}}$ that is also used for lowering and raising indices, and field strength (torsion) corresponding to the $B$-field  is defined by
\begin{eqnarray}\label{3.25}
H_{_{MNP}} =(-1)^{^{M}}\;\frac{\overrightarrow{\partial}}{\partial
X^{^{M}}} B_{_{NP}}+(-1)^{^{N+M(N+P)}}\;\frac{\overrightarrow{\partial}}{\partial
X^{^{N}}} B_{_{PM}}+(-1)^{^{P(1+M+N)}}\;\frac{\overrightarrow{\partial}}{\partial
X^{^{P}}} B_{_{MN}}.~~~~
\end{eqnarray}
Now we want to obtain conditions under which the background given by \eqref{3.20} and \eqref{3.21} satisfies the beta-function equations \eqref{3.22}-\eqref{3.24}.
These equations possess a solution with the metric \eqref{3.20} and $B$-field \eqref{3.21} and also a constant dilaton field
$\Phi = \phi_{_0}$ if the following conditions hold:
\begin{eqnarray}\label{3.26}
F(r) &=& F_{_0},~~~~~~~ k(r) = { r^2\over l},~~~~~~~ h_2(r) = 0, \nonumber\\
h_1(r) &=&
\begin{cases}
\frac{-{\cal C}_0}{8(r^2-\frac{1}{2}Ml^2)}~~~~~~ &~~~~~~ J=Ml,~~~~~\\
\frac{{\cal C}_0}{4l\sqrt{M^2 l^2 - J^2}} \tanh^{-1} \Big(\frac{Ml^2 - 2 r^2}{l\sqrt{M^2 l^2 - J^2}}\Big) ~~~~~~&~~~~~~ J^2 < M^2 l^2,\\
{{{\cal C}_0} \over 4l\sqrt{J^2 - M^2 l^2}} \tan^{-1} \Big(\frac{ 2 r^2 - Ml^2}{l\sqrt{J^2 - M^2 l^2}}\Big)~~~~~~ &~~~~~~ J^2>M^2 l^2,
\end{cases}
\end{eqnarray}
for some constant ${{\cal C}_0}$. Notice that here the cosmological constant is obtained to be
$\Lambda = -1/{l^2}$.
Taking into account the above results, the scalar curvature of the metric \eqref{3.20} is ${\cal R} =-6/l^2$,
and this is exactly the same as the case where fermion fields are absent. Furthermore,
for the $B$-field, the corresponding field strength is given by
\begin{eqnarray}\label{3.27}
H= -\frac{2 r}{l}~ dr \wedge dt \wedge d\phi + H_{_{144}}~ dr \wedge d\psi \wedge d\psi,
\end{eqnarray}
where
\begin{eqnarray}\label{3.28}
H_{_{144}} =
\begin{cases}
\frac{C_0 r}{(2 r^2 -Ml^2)^2}~~~~~~ &~~~~~~ J=Ml,~~~~~\\
\frac{C_0 r}{J^2 l^2-4Ml^2 r^2+4r^4}~~~~~~&~~~~~~ J^2 < M^2 l^2, ~J^2>M^2 l^2.\\
\end{cases}
\end{eqnarray}

\subsubsection{The dual model}
In the same way, to construct out the dual $\sigma$-model on the supermanifold
$\tilde {{\M}} \approx O \times \tilde {G}$
we parameterize  Abelian Lie supergroup  $\tilde G$ with the bosonic coordinates
$(\tilde t, \tilde \varphi)$  and fermionic ones $(\tilde \psi, \tilde \chi)$ so that its element
is defined as \eqref{3.15} by replacing untilded quantities with tilded ones.
In order to obtain the dual $\sigma$-model to \eqref{3.17}, we use the action
\eqref{3.7}. By taking into account \eqref{3.26} for \eqref{3.16} and then by
inserting the result into \eqref{3.11} one can obtain
the dual coupling matrices. They are then read
\begin{eqnarray}
{\tilde E}^{{ab}} &=&\left( \begin{tabular}{cccc}
                 $\frac{4r^2}{4Mr^2-J^2}$ & $\frac{2Jl+4r^2}{l(4Mr^2-J^2)}$ & 0 & 0 \\
                 $\frac{2Jl-4r^2}{l(4Mr^2-J^2)}$ & $\frac{4(Ml^2-r^2)}{l^2(4Mr^2-J^2)}$  & 0& 0 \\
                 0 & 0& $0$ & $\frac{-1}{F_0}$ \\
                 0 & 0 & $\frac{1}{F_{_0}}$ & $\frac{h_1(r)}{F_{_0}^2}$ \\
                 \end{tabular} \right),\nonumber\\
{\tilde \phi}_{_{ij}}  &=&  \big({  r^2\over l^2} - {M}  +
{J^2\over 4r^2}\big)^{-1},~~~~~~~~~{\tilde \phi}^{\hspace{0mm}{(1)^{ a}}}_{~~~j}  = {\tilde \phi}^{\hspace{0mm}{(2)^{ b}}}_{i} = 0,\label{3.29}
\end{eqnarray}
where $h_1(r)$ is given by relation \eqref{3.26}.
Finally, by using \eqref{3.29} and \eqref{3.7} one can construct out the action of dual model.
The line element and anti-supersymmetric field corresponding
to the dual action may be expressed as
\begin{eqnarray}
d{\tilde s}^2 &=& \big({ r^2\over l^2} - M  +
{J^2\over 4r^2}\big)^{-1} dr^2 + \big(\frac{1}{Mr^2-\frac{J^2}{4}}\big)\Big[r^2 d\tilde t^2 + J d\tilde t d\tilde \phi \nonumber\\
&& ~~~~~~~~~~~~~~~~~~~~~~~~~~~~~~~~~+(M-\frac{r^2}{l^2}) d\tilde \phi^2\Big] - \frac{2}{F_0}~ d\tilde \psi d\tilde \chi,\nonumber\\
{\tilde B }&=& \frac{{4 r^2}}{l(4Mr^2-{J^2})}~ d\tilde t \land d\tilde \phi  - { h_1(r)\over{2F_0^2}}~ d\tilde \chi \land d\tilde \chi.\label{3.30}
\end{eqnarray}
The components of metric are ill defined at the region $r = \frac{J}{2} M^{\frac{-1}{2}}$.
We can test whether there is true singularity
by calculating the scalar curvature, which is
\begin{eqnarray}
{\tilde {\cal R}} ~=~ \frac{16 M^2 l^2(4Mr^2-{J^2})-2J^2 (3J^2 +16 Mr^2)}{l^2(4Mr^2-{J^2})^2}.\label{3.31}
\end{eqnarray}
It is clear that $r = \frac{J}{2} M^{\frac{-1}{2}}$ is a true curvature singularity,
so that it cannot be removed by a change of coordinates. Moreover,
the horizons of the dual metric are at the same location as the metric of original model.
We also find that the field strength corresponding to the $\tilde B$-field is
\begin{eqnarray}\label{3.32}
\tilde H= \frac{-8 rJ^2}{l (J^2-4Mr^2)^2}~ dr \wedge d\tilde t \wedge d\tilde \phi  + \frac{H_{_{144}}}{F_{_0}^2}~ dr \wedge d\tilde \chi \wedge d\tilde \chi,
\end{eqnarray}
where $H_{_{144}}$ is given by relation \eqref{3.28}. The dual dilaton field that makes the model conformal is obtained by
using equation \eqref{3.21.1} to be
\begin{eqnarray}\label{3.33}
\tilde \Phi =\phi_0-{1\over2} \log({Mr^2-\frac{J^2}{4}}).
\end{eqnarray}
Thus, the dual metric with the field strength \eqref{3.32} and dilaton field \eqref{3.33}
satisfy one-loop beta-function equations \eqref{3.22}-\eqref{3.24}
in a way that the cosmological constant must be $\tilde \Lambda = -{1}/{l^2}$.


\subsection{Abelian T-duality with Abelian Lie supergroup of the type $(1|2)$}

In order to study Abelian T-duality of the BTZ black hole metric coupled to two fermionic fields,
there is a possibility that we can perform the dualizing on Abelian Lie supergroups $G$ and $\tilde G$ of the type $(1|2)$; moreover,
the spectator fields are considered to be
$y^i =(r, t)$. As we will see, the original model will be the same as \eqref{3.17}.
Here we find a new Abelian target space dual
for the action \eqref{3.17}. In fact, the result of duality will be different from those of
\eqref{3.30} and also \eqref{3.31}.

In order to construct out the original $\sigma$-model, we parameterize an element of $G$ as
\begin{eqnarray}\label{3.34}
g= e^{\varphi T_{_1}} e^{\psi T_{_2}} e^{\chi T_{_3}},
\end{eqnarray}
where $\varphi$ is only the bosonic coordinate, while  $(\psi, \chi)$ denote fermionic coordinates.
Also, $\{T_{_1}\}$ and $\{T_{_2},  T_{_3}\}$ are bosonic and fermionic generators of $G$, respectively.
In this case, the dualizing is performed on the coordinates $(\varphi; \psi, \chi)$, so
it is more appropriate to choose the spectator-dependent matrices as follows:
\begin{eqnarray}\label{3.35}
{E_{0}}_{_{ab}}&=& \left( \begin{tabular}{ccc}
                 $r^2$ & 0 & 0 \\
                  0& $h_1(r)$ & $F_0$ \\
                  0 & -$F_0$ & $0$ \\
                 \end{tabular} \right),~~~~~F^{(1)}_{a i}  = \left( \begin{tabular}{cc}
                0 & $-\frac{J}{2}+\frac{r^2}{l}$ \\
                $0$ & $0$ \\
                $0$ & $0$ \\
                 \end{tabular} \right),\nonumber\\
F^{(2)}_{i a} &=& \left( \begin{tabular}{ccc}
                  0& $0$ & $0$ \\
                $-\frac{J}{2}-\frac{r^2}{l}$ & $0$ & $0$ \\
                 \end{tabular} \right),~~~~~~F_{i j}  = \left( \begin{tabular}{cc}
                 $(\frac{r^2}{l^2}-M+\frac{J^2}{4r^2})^{-1}$ & $0$ \\
                0  & $M-\frac{r^2}{l^2}$ \\
                 \end{tabular} \right).
\end{eqnarray}
where $h_1(r)$ is given by relation \eqref{3.26}.
Then, using the fact that $\Pi(g)=0$ and utilizing formulae \eqref{3.8}
together with \eqref{3.4}, the original $\sigma$-model is obtained to be the same as \eqref{3.17}
provided that one also considers relation \eqref{3.26}.

The corresponding coupling matrices to the dual model can be obtained by making use of relations \eqref{3.11} and \eqref{3.35}.
They are then read
\begin{eqnarray}\label{3.36}
{\tilde E}^{{ab}} &=&\left( \begin{tabular}{ccc}
                 $\frac{1}{r^2}$ & 0 & 0 \\
                 0 & 0& $\frac{1}{F_0}$ \\
                 0 & $\frac{-1}{F_0}$ & $\frac{-h_1(r)}{F_0^2}$ \\
                 \end{tabular} \right),~~~~~~{\tilde \phi}^{\hspace{0mm}{(1)^{ a}}}_{~~~j}  = \left( \begin{tabular}{cc}
                0 & $\frac{-J}{2r^2}+\frac{1}{l}$ \\
                $0$ & $0$ \\
                $0$ & $0$ \\
                 \end{tabular} \right),\nonumber\\
{\tilde \phi}^{\hspace{0mm}{(2)^{ b}}}_{i} &=& \left( \begin{tabular}{ccc}
                  0& $0$ & $0$ \\
                $\frac{J}{2r^2}+\frac{1}{l}$ & $0$ & $0$ \\
                 \end{tabular} \right),~~~~~~~{\tilde \phi}_{_{ij}}  = \left( \begin{tabular}{cc}
                 $(\frac{r^2}{l^2}-M+\frac{J^2}{4r^2})^{-1}$ & $0$ \\
                0  & $M-\frac{J^2}{4r^2}$ \\
                 \end{tabular} \right).\label{3.36}
\end{eqnarray}
Thus, employing \eqref{3.7} the dual background can be cast in the form
\begin{eqnarray}\label{3.37}
d\tilde s^2 &=& \big({  r^2\over l^2} - { M}  +
{J^2\over 4r^2}\big)^{-1} dr^2 + (M-\frac{J^2}{4r^2}) dt^2 + {2\over l} dt d\tilde \phi
+{1\over r^2} d\tilde \phi^2 - \frac{2}{F_0} d\tilde \psi d\tilde \chi,\nonumber\\
\tilde B &=& \frac{J}{2r^2}~ dt \land d\tilde \phi  + { h_1(r)\over{2F_0^2}}~ d\tilde \chi \land d\tilde \chi,
\end{eqnarray}
where $h_1(r)$ is given by relation \eqref{3.26}. Let us now enhance and clarify the structure of the dual spacetime.
As it is seen, the dual metric has an apparent singularity.
Calculating the scalar curvature, which is
\begin{eqnarray}
{\tilde {\cal R}} ~=~ \frac{8 M r^2-7 J^2}{2 r^4},\label{3.38}
\end{eqnarray}
one should test whether there is a true singularity, and we thus see that $ r=0 $ is a true singularity.
In addition to the above, one immediately verifies the field equations \eqref{3.22}-\eqref{3.24}
for the dual background with the dilaton
field $\tilde \Phi =\phi_{_0}- \log r$;
moreover, to satisfy the equations we must have a cosmological constant as in the original model, i.e.,
$\tilde \Lambda = -{1}/{l^2}$. Notice that the dual dilaton field satisfies formula \eqref{3.21.1}.

In summary, we obtained two Abelian duals for the BTZ metric coupled to two fermionic fields.
When the dualizing was implemented by the Abelian Lie supergroup of the type $(2|2)$, we found that the singularity of the dual metric
has appeared at the point $r = \frac{J}{2} M^{\frac{-1}{2}}$,
whereas when we deal with a $(1|2)$-dimensional Abelian Lie supergroup, we encounter a singularity at the origin.
Moreover, in both cases of the dual metrics, horizons are at the same location as the metric of original model.
\vspace{-7mm}
\section{Non-Abelian T-duality of BTZ vacuum metric coupled to two fermionic fields}
\label{Sec.IV}
\vspace{-3mm}
Our goal in this section is that to calculate non-Abelian target space duals (here as super PL T-duality on a semi-Abelian superdouble)
of the BTZ vacuum metric when is coupled to two fermionic fields.
The Lie supergroup acting freely on the original target supermanifold of $T$-dual $\sigma$-models is considered to be one of the
non-Abelian Lie supergroups of the type $(2|2)$ whose Lie superalgebras were classified by Backhouse in \cite{B} as mentioned in Introduction.
The classification of decomposable and indecomposable Lie superalgebras
of the type $(2|2)$ has been displayed into six disjoint families in Tables 2 to 7 of Appendix A.
In this section we perform the duality on some Lie supergroups whose Lie superalgebras are considered to be the $({\C}_1^1 +{\A})$ of Family $I$,
$({\C}^3 +{\A})$ of Family $II$, ${\C}^3 \oplus {\A}_{1,1}$ of Family $III$,
${\D}^{10}_{p=\pm\frac{1}{2}}$ of Family $IV$ and $(2{ \A}_{1,1}+2{ \A})^0$ of Family $V$.
These Lie superalgebras possess two bosonic generators $\{T_{_1}, T_{_2}\}$ along with two
fermionic ones $\{T_{_3}, T_{_4}\}$\footnote{From now on we consider $\{T_{_1}, T_{_2}\}$ and $\{T_{_3}, T_{_4}\}$ as bosonic and fermionic generators, respectively.}.
Note that in order to satisfy the dualizability conditions, the Lie supergroup of dual target supermanifold must be chosen Abelian.
In fact, we are dealing with semi-Abelian superdoubles which are non-isomorphic as Lie superalgebras in each of the models.
\vspace{-4mm}
\subsection{Non-Abelian target space dual starting from the $(C_1^1 +A)$ Lie supergroup}
In what follows we shall obtain the BTZ vacuum metric coupled to two fermionic fields from a $T$-dualizable $\sigma$-model
constructing on a five-dimensional target supermanifold ${\M} \approx O \times  G$ where $ G$ as the first sub-supergroup of
Drinfeld superdouble is considered to be the $(C_1^1 +A)$
acting freely on ${\M}$, while $O$ as the orbit of ${ G}$ in ${\M}$ is a one-dimensional space with the coordinate $y^i = \{y\}$.
The second sub-supergroup, ${\tilde { G}} = {I}_{_{(2|2)}}$, acting freely
on the dual supermanifold ${\tilde {\M}} \approx O \times {\tilde { G}}$ is assumed to be Abelian of the type $(2|2)$.
Hence, the super PL T-duality reduces to the super non-Abelian T-duality.
One can find the (anti-)commutation relations of the $({\C}^1_1+{\A})$ Lie superalgebra in Table 2 of Appendix A.
As mentioned in Introduction section,  having  Drinfeld superdoubles one can construct super PL T-dual $\sigma$-models on them.
The Lie superalgebra of the Drinfeld superdouble which we refer to as
$\big(({\C}^1_1+{\A})\;,\;{\cal I}_{_{(2|2)}}\big)$ is defined by the following nonzero (anti-)commutation relations\footnote{Notice that
in Ref. \cite{Eghbali}, the equations of motion of a super non-Abelian T-dual $\sigma$-model on the  $(C_1^1 +A)$ Lie
supergroup in the curved background have been explicitly solved by
making use of the transformation between the supergroup coordinates of the model
living in the flat background and its flat supercoordinates.}:
\begin{eqnarray}\label{3.39}
{[T_{_1} , T_{_2}]} &=& T_{_2},~~~~~~~~[T_{_1} , T_{_3}]=T_{_3},~~~~~~~\{T_{_3} , T_{_4}\}=T_{_2},\nonumber\\
{[T_{_1} , {\tilde T}^2]}&=&-{\tilde T}^2,~~~~~[T_{_1} , {\tilde T}^3]=-{\tilde T}^3,~~~~[T_{_2} , {\tilde T}^2]={\tilde T}^1,\nonumber\\
{[T_{_3} , {\tilde T}^2]} &=& -{\tilde T}^4,~~~~~[T_{_4} , {\tilde T}^2]=-{\tilde T}^3,~~~~\{T_{_3} , {\tilde T}^3\}=-{\tilde T}^1,
\end{eqnarray}
where $\{T_{_1}, T_{_2}; T_{_3}, T_{_4}\}$ generate the $({\C}^1_1+{\A})$, and $\{ {\tilde T}^1, {\tilde T}^2\}$ and $\{{\tilde T}^3 , {\tilde T}^4\}$ are the respective bosonic and fermionic generators of the
${\cal I}_{_{(2|2)}}$. In order to calculate the components of the right-invariant super
one-forms $R_{\pm}^a$ on the $(C_1^1 +A)$ Lie supergroup we parameterize an element of the supergroup as
\begin{eqnarray}\label{3.40}
g~=~e^{x_{_1} T_{_1}}~e^{x_{_2} T_{_2}} ~e^{\psi T_{_3}}~e^{\chi T_{_4}},
\end{eqnarray}
where $(x_{_1}, x_{_2})$ are bosonic fields, while $(\psi, \xi)$ stand for fermionic ones.
From now on, we will use this notation for bosonic and fermionic fields. Now, one may use equation \eqref{3.5} to get
\begin{eqnarray}\label{3.41}
R_{\pm}^1&=& \partial_{\pm} x_{_{1}},~~~~~~~~~~~~~~~R_{\pm}^2 ~=~ \partial_{\pm} x_{_{2}}  e^{x_{_{1}}} + \partial_{\pm} \chi ~ \psi e^{x_{_{1}}},\nonumber\\
R_{\pm}^3&=& -\partial_{\pm} \psi~ e^{x_{_{1}}}, ~~~~ ~~~~~R_{\pm}^4 ~=~ -\partial_{\pm} \chi.
\end{eqnarray}
Let us now choose the spectator-dependent background matrices as
\begin{eqnarray}
{E_{0}}_{_{ab}}=\left( \begin{tabular}{cccc}
                 $0$ & $\frac{1}{2} e^{-2 y}$ & 0 & 0 \\
                 $\frac{1}{2} e^{-2 y}$ & $0$  & 0& 0 \\
                 0 & 0& $0$ & $h(y)$ \\
                 0 & 0 & -$h(y)$ & $\alpha_{_0} n(y)$ \\
                 \end{tabular} \right),~~~~~F^{(1)}_{a i}  = F^{(2)}_{i a} = 0,~~~~F_{i j} = l^2,\label{3.42}
\end{eqnarray}
where $\alpha_{_0} $ is an arbitrary constant. Also, $h(y)$ and $n(y)$ are functions of the variable $\{y\}$ only.
In the case of the super non-Abelian T-duality,
it can be deduced from formula \eqref{3.10} that the Poisson superbracket $\Pi(g)$ vanishes provided that the
dual Lie supergroup is Abelian. Thus, by using these and also employing \eqref{3.4} one can get the action of original model
such that its corresponding line element and $B$-field are given by
\begin{eqnarray}
ds^2&=& l^2  dy^2 + e^{{x_{_{1}}} -2y} ~ d x_{_{1}}  d x_{_{2}} - \psi   e^{{x_{_{1}}} -2y}~ d x_{_{1}}  d\chi
- 2 h(y) e^{x_{_{1}}} ~ d\psi d\chi,\label{3.43}\\
B&=& -\frac{1}{2} \alpha_{_0} n(y)~ d\chi \wedge d\chi.\label{3.44}
\end{eqnarray}
The scalar curvature of the metric is
\begin{eqnarray}
{\cal R} = -\frac{6} {l^2} - \frac{1} {2 l^2 h^2(y)} \Big[ 8h(y) h'(y) +5 {h'}^2(y)-4 h(y) h''(y) \Big],\label{3.45}
\end{eqnarray}
where prime denotes differentiation with respect to the argument $\{y\}$.
It is straightforward to verify that the only non-zero component of the field strength corresponding to $B$-field
\eqref{3.44} is $H{_{y \chi \chi}}=\alpha_{_0} n'(y)$.
We are now interested in the satisfaction of the conformal invariance conditions of the model.
In order to guarantee the
conformal invariance of the model, at least at the one-loop level,
one must show that the background of model including \eqref{3.43} and \eqref{3.44} satisfies the vanishing of the beta-function equations.
Before we proceed to investigate these conditions further, let us briefly comment on the
formula of dilaton transformation for the case of PL T-duality.
In \cite{vonUnge}, von Unge showed that the duality transformation must be supplemented by a
correction that comes from integrating out the fields on the dual group
in path-integral formulation so that it can be absorbed at the one-loop level into the transformation of the dilaton field.
Following \cite{vonUnge}, in the super PL T-duality case, these transformations may be expressed as
\begin{eqnarray}
\Phi &=&\phi^{^{(0)}} +\frac{1}{2}\log \big|s\hspace{-0.6mm}\det\big({_{_a}}{E_{_b}}(g, y^i)\big) \big| -\frac{1}{2}\log \big|s\hspace{-0.6mm}\det\big({_{_a}}{{E_0}_{_b}}(y^i)\big)\big| -\frac{1}{2}\log \big|s\hspace{-0.6mm}\det\big({_{_a}}{a^{^{b}}} {\tiny (g)}\big)\big|,~~~~~~\label{3.45.1}\\
{\tilde \Phi} &=& \phi^{^{(0)}}+\frac{1}{2} \log |s\hspace{-0.6mm}\det\big({{{\tilde E}^{ab}}}(\tilde g, y^i)\big)| - \frac{1}{2} \log|s\hspace{-0.6mm}\det\big({{{\tilde a}^a}_{_{~}b} ({\tilde g})}\big)|,\label{3.45.2}
\end{eqnarray}
where $\phi^{^{(0)}}$ may depend on both supergroup and spectator coordinates.
Note that in the case of super non-Abelian T-duality, formula \eqref{3.45.1} becomes simpler, because we have $E= {E_0}$.
In the present example, sub-matrix $a_{_{a}}^{^{~b}}(g)$ can be obtained from the first equation of \eqref{3.10}, then, one can easily find that
$s\hspace{-0.6mm}\det\big({_{_a}}{a^{^{^b}}} {\tiny (g)}\big)=1$. Consequently, it follows from  \eqref{3.45.1} that $\Phi =\phi^{^{(0)}}$
such that the beta-function equations
are satisfied with a constant dilaton field, $\Phi = \phi^{^{(0)}}={\cal C}_{_0}$,
and vanishing cosmological constant if functions $h(y)$ and $n(y)$ obey the following relations
\begin{eqnarray}
h(y) =  -\frac{1}{2} e^{-2 y},~~~~~~~~~~~ n(y) =  1-\frac{1}{4} e^{-4 y}.\label{3.46}
\end{eqnarray}
It can be useful to comment on the fact that the metric \eqref{3.43} along with $h(y)$ obtained in
\eqref{3.46} is flat in the sense that its scalar curvature and Ricci tensor vanish. On the other hand,
one may use the coordinate transformation
\begin{eqnarray}
e^{y} = \frac{l}{r},~~~~~~~~~~~ e^{x_{_{1}}} =  \frac{(t- l \varphi)}{l},~~~~~~~~~ {x_{_{2}}}= -l (t+l \varphi),\label{3.47}
\end{eqnarray}
then, the background of model, that is conformally invariant up to one-loop order, reduces to
\begin{eqnarray}
ds^2&=& \frac{l^2}{r^2}  dr^2 - \frac{r^2}{l^2} dt^2 +r^2 d\varphi^2 + \frac{r^2}{l^2}\big(-\frac{\psi}{l} dt d\chi +\psi
d\varphi d\chi + d\psi d\chi\big), \nonumber\\
H&=& -\frac{\alpha_{_0} r^3}{2 l^4}~ dr \wedge d\chi \wedge d\chi,\nonumber\\
\Phi &=& {\cal C}_{_0}. \label{3.48}
\end{eqnarray}
Indeed, the above metric is nothing but the BTZ vacuum metric which has been
coupled to two fermionic fields $(\psi, \chi)$. A remarkable point about the metric in \eqref{3.48} is that
unlike $2+1$-dimensional BTZ vacuum metric (the $J=M=0$ case of \eqref{int.1})
which has a constant negative scalar curvature, it has vanishing scalar curvature.
In fact, by coupling the fermionic fields to the metric, the scalar curvature undergoes a change.
It is then straightforward to compute the corresponding dual background. By considering the coupling matrices in
\eqref{3.42} and by employing \eqref{3.11} one can show that the dual couplings take the following form
\begin{eqnarray}
{{\tilde E}^{^{ab}}} = {\big(E_{0}+ {\tilde \Pi}\big)^{\hspace{-1mm}-1}}^{ab},~~~~~~
{\tilde \phi}^{\hspace{0mm}{(1)^{ a}}}_{~~~j} = {\tilde \phi}^{\hspace{0mm}{(2)^{ b}}}_{i} = 0,~~~~~~~~
{\tilde \phi}_{_{ij}} = F_{_{ij}}=l^2,\label{3.49}
\end{eqnarray}
in which the Poisson superbracket ${\tilde \Pi}_{ab}$ can be obtained from equation \eqref{3.10} by replacing untilded quantities with tilded ones.
To this end,  we  parameterize the dual Lie supergroup ${I}_{_{(2|2)}}$ with
coordinates ${\tilde x}_a=({\tilde x}_{_1}, {\tilde x}_{_2}; {\tilde \psi}, {\tilde \xi})$ so that its elements
are defined as in \eqref{3.40}. We then find
\begin{eqnarray}
{\tilde \Pi}_{ab}=\left( \begin{tabular}{cccc}
                 $0$ & -${\tilde x}_{_2}$ & -${\tilde \psi}$ & 0 \\
                 ${\tilde x}_{_2}$ & $0$  & 0& 0 \\
                 ${\tilde \psi}$ & 0& $0$ & -${\tilde x}_{_2}$ \\
                 0 & 0 & -${\tilde x}_{_2}$ & 0 \\
                 \end{tabular} \right).\label{3.50}
\end{eqnarray}
Putting \eqref{3.50} and $E_{0}$ from \eqref{3.42} together into \eqref{3.49} and using the
fact that the components of the right-invariant super one-forms on
${I}_{_{(2|2)}}$ are ${\tilde R}_{\pm_{a}}=\partial_{\pm} {{\tilde x}_a}$, the background of dual $\sigma$-model is obtained to be
\begin{eqnarray}
{\tilde ds^2}&=& {l^2}  dy^2 + \frac{1}{\Delta}\Big[-e^{-2y}  d{\tilde x}_{_1} d {\tilde x}_{_2} + \frac{\alpha_{_0}}{\Delta}  (1-\frac{1}{4} e^{-4 y}) {\tilde \psi} d {\tilde x}_{_2} d {\tilde \psi} \nonumber\\
&&~~~~~~~~~~~~~~~~~~~~~~~~~~~~~~~~~~+ \frac{2}{\Delta} e^{-2y} {\tilde x}_{_2}  {\tilde \psi} d {\tilde x}_{_2} d {\tilde \chi}- e^{-2y} ~d {\tilde \psi}  d {\tilde \chi}\Big],\label{3.51}\\
{\tilde B}&=& \frac{1}{\Delta}\Big[{\tilde x}_{_2}  ~ d{\tilde x}_{_1}  \wedge d {\tilde x}_{_2} -{\tilde x}_{_2} ~d{\tilde \psi} \wedge  d {\tilde \chi}-\frac{\alpha_{_0}}{2} (1-\frac{1}{4} e^{-4 y})  d{\tilde \psi} \wedge d{\tilde \psi}\nonumber\\
&&~~~~~~~~+\frac{\alpha_{_0}}{\Delta}  (1-\frac{1}{4} e^{-4 y}) {\tilde x}_{_2} {\tilde \psi} d {\tilde x}_{_2} \wedge  d {\tilde \psi}
+\frac{1}{\Delta} ({{\tilde x}_{_2}}^2 +\frac{1}{4} e^{-4 y}) {\tilde \psi} d {\tilde x}_{_2} \wedge d {\tilde \chi}\Big],\label{3.52}
\end{eqnarray}
where $\Delta ={{\tilde x}_{_2}}^2 -\frac{1}{4} e^{-4y} $. The metric \eqref{3.51} is flat in the sense that its scalar
curvature vanishes. Looking at the field equations \eqref{3.22}-\eqref{3.24}
one verifies the conformal invariance of dual model with a constant dilaton field and vanishing cosmological constant.
Notice that the dilaton field making the dual model conformal can be also obtained
from \eqref{3.45.2} which gives a constant value.
Finally, it is seen that in this case, the super PL T-duality transforms
rather extensive and complicated background given by equations \eqref{3.51} and \eqref{3.52} to much simpler form such as \eqref{3.48}.

Before closing this subsection let us highlight some point concerning these models.
As discussed in Introduction, the BTZ metric \eqref{int.1} (even for the vacuum state) satisfies the one-loop beta-function equations with a non-vanishing
field strength, while here, as can be clearly seen, the original model with background \eqref{3.48} satisfies the one-loop beta-function equations
in both cases of absence $(\alpha_{_0}=0)$ and presence $(\alpha_{_0}\neq 0)$ of the field strength.

\subsection{Non-Abelian target space dual starting from the $(C^3 +A)$ Lie supergroup}

The $({\C}^3 + {\A})$ Lie superalgebra\footnote{In Ref. \cite{ER8}, it has been shown that the super PL duality relates the
 WZW model based on the $(C^3+A)$  Lie
supergroup to a $\sigma$-model defined on the $(C^3+A)$  when Lie superalgebra  of the dual Lie supergroup is  ${{\C}^3 \oplus
{\A}_{_{1,1}}}$.} is in turn interesting in the sense that its commutator of B-B is zero.
Its (anti-)commutation relations are given in Table 3 of Appendix A.
The Lie superalgebra of semie-Abelian Drinfeld superdouble
$\big(({\C}^3 + {\A})\;,\;{\cal I}_{_{(2|2)}}\big)$ obeys the following (anti-)commutation relations
\begin{eqnarray}\label{3.53}
{[T_{_1} , T_{_4}]} &=& T_{_3},~~~~~~~~~~~~~~\{T_{_4} , T_{_4}\}=T_{_2},~~~~~~~~~~~\{T_{_1} , {\tilde T}^3\}=-{\tilde T}^4,\nonumber\\
{[T_{_4} , {\tilde T}^2]} &=& -{\tilde T}^4,~~~~~~~~~~~\{T_{_4} , {\tilde T}^3\}=-{\tilde T}^1,
\end{eqnarray}
where $T_{_a}$'s generate the $({\C}^3+{\A})$,
while ${\tilde T}^a$'s stand for the ${\cal I}_{_{(2|2)}}$.
The expression for a generic element of the $(C^3 +A)$ Lie supergroup  can
be written as
\begin{eqnarray}\label{3.54}
g~=~ e^{\psi T_{_3}} ~ e^{x_{_1} T_{_1}}~e^{x_{_2} T_{_2}} ~e^{\chi T_{_4}}.
\end{eqnarray}
In order to construct the original $\sigma$-model we need to calculate the right-invariant one-forms on the $(C^3 +A)$.
To this end, one should employ \eqref{3.53} and \eqref{3.54} and then use equation \eqref{3.5}. The result is
\begin{eqnarray}\label{3.55}
R_{\pm}^1&=& \partial_{\pm} x_{_{1}},~~~~~~~~~~~~~~~~~~~~~~~~~R_{\pm}^2 ~=~ \partial_{\pm} x_{_{2}}
+ \partial_{\pm} \chi ~ \frac{\chi}{2},\nonumber\\
R_{\pm}^3&=& -\partial_{\pm} \psi- \partial_{\pm} \chi~  x_{_{1}}, ~~~~~~~~~~R_{\pm}^4 ~=~ -\partial_{\pm} \chi.
\end{eqnarray}
In what follows, the dualizing is performed on the $(C^3 +A)$ Lie supergroup, and the spectator field is still displayed by $\{y\}$.
Here we choose the spectator-dependent background matrices in the following form
\begin{eqnarray}
{E_{0}}_{_{ab}}=\left( \begin{tabular}{cccc}
                 $a_{_1} (y)$ & 0 & 0 & 0 \\
                 0 & $a_{_2} (y)$  & 0& 0 \\
                 0 & 0& $0$ & $a_{_3} (y)$ \\
                 0 & 0 & -$a_{_3} (y)$ & $b(y)$ \\
                 \end{tabular} \right),~~~~~F^{(1)}_{a i}  = F^{(2)}_{i a} = 0,~~~~F_{i j} = \frac{l^2}{4},\label{3.56}
\end{eqnarray}
where $a_{_i} (y),~i=1, 2, 3$ and $b(y)$ are some functions
that may depend on the coordinate $\{y\}$. Now, by applying formulae \eqref{3.4} and \eqref{3.8} one
can obtain the background of the original $\sigma$-model, giving us
\begin{eqnarray}
ds^2&=& \frac{l^2}{4}  dy^2 + a_{_1} (y) ~ {d x_{_{1}}}^2  + a_{_2} (y) ~ {d x_{_{2}}}^2 - a_{_2} (y)  \chi d x_{_{2}} d \chi-2 a_{_3} (y) ~ d\psi d\chi,\label{3.57}\\
B&=& -\frac{1}{2} b(y)~ d\chi \wedge d\chi.\label{3.58}
\end{eqnarray}
In order to obtain dilaton field
that supports the background of the model one may apply formula \eqref{3.45.1}. First, by using the first equation of \eqref{3.10} and
(anti-)commutation relations of the $({\C}^3 + {\A})$ given in \eqref{3.53}
one immediately concludes that $s\hspace{-0.6mm}\det\big({_{_a}}{a^{^{^b}}} {\tiny (g)}\big)=1$. Hence,
from  \eqref{3.45.1} we find that $\Phi =\phi^{^{(0)}}$. As we will see, the dilaton field that makes the original $\sigma$-model conformal
is found to be $\Phi = {\cal C}_{_0} + y$, so we choose $\phi^{^{(0)}} = {\cal C}_{_0} + y$.
Looking at the one-loop conformal invariance conditions, the field equations \eqref{3.22}-\eqref{3.24} are
then satisfied with the metric \eqref{3.57}, $B$-field \eqref{3.58} and a zero cosmological constant
together with the aforementioned dialton field if one considers
\begin{eqnarray}
a_{_1} (y)= -a_{_2} (y)= e^y,~~~~~~~a_{_3} (y)= e^{-y},~~~~~~b(y) =\beta_{_0} (1-\frac{1}{2} e^{-2y}),\label{3.59}
\end{eqnarray}
for some constant $\beta_{_0}$.
Inserting \eqref{3.59} into \eqref{3.57} one immediately finds that the scalar
curvature of the metric is non-zero, and it has a negative constant value as ${\cal R} =-16/l^2$.
It can be easily shown that under the coordinate transformation
\begin{eqnarray}
e^{y} = {r}^2,~~~~~~~~~~~ {x_{_{1}}} = \varphi,~~~~~~~~~ {x_{_{2}}}= \frac{1}{l} t,   \label{3.60}
\end{eqnarray}
the background becomes
\begin{eqnarray}
ds^2&=& \frac{l^2}{r^2}  dr^2 - \frac{r^2}{l^2} dt^2 +r^2 d\varphi^2 + \frac{r^2}{l} \chi dt d\chi -\frac{2}{r^2} d\psi d\chi, \nonumber\\
H&=& -\frac{\beta_{_0}}{ r^5}~ dr \wedge d\chi \wedge d\chi,\nonumber\\
\Phi &=& {\cal C}_{_0} +2 \log r. \label{3.61}
\end{eqnarray}
Again, the above metric is nothing but the BTZ vacuum metric
coupling to two fermionic fields. Note that here the terms related to the coupling of fermions to the metric are different from those of metric in
\eqref{3.48}. Analogously, we here can have a conformal background in both cases of absence $(\beta_{_0}=0)$ and presence ($\beta_{_0}\neq 0$)
of the field strength.

To continue, we obtain a new dual background for the BTZ vacuum metric coupled to two fermionic fields.
Employing \eqref{3.11}, \eqref{3.56} and \eqref{3.59} one can get the dual couplings.
In this case, the Poisson superbracket and non-zero background matrices read
\begin{eqnarray}
{\tilde \Pi}_{ab}=\left( \begin{tabular}{cccc}
                 0 & 0 & 0 & -$ {\tilde \psi}$ \\
                 0 & $0$  & 0& 0 \\
                 0 & 0& $0$ & 0 \\
                 $ {\tilde \psi}$ & 0 & 0 & -${\tilde x}_{_2}$ \\
                 \end{tabular} \right),~~{\tilde E}^{{ab}} =\left( \begin{tabular}{cccc}
                 $e^{-y}$ & 0 & -$ {\tilde \psi}$ & 0 \\
                 0 & -$e^{-y}$  & 0 & 0 \\
                 $ {\tilde \psi}$  & 0 & $ {\tilde x}_{_{2}} e^{2y}$ & $e^{y}$ \\
                 0 & 0 & -$e^{y}$ & 0 \\
                 \end{tabular} \right),~~{\tilde \phi}_{_{ij}} =\frac{l^2}{4},~~~\label{3.62}
\end{eqnarray}
where we have here assumed that $\beta_{_0}=0$.
Finally, with the help of \eqref{3.7} the line element and the tensor
field ${\tilde B}$ take the following forms
\begin{eqnarray}
{\tilde ds^2}&=& \frac{l^2}{4}  dy^2 + e^{-y}  \big({d {\tilde x}_{_{1}}}^2  - {d {\tilde x}_{_{2}}}^2\big) - 2 e^{y} ~ d {\tilde \psi} d{\tilde \chi},\label{3.63}\\
{\tilde B}&=& {\tilde \psi} ~  {d {\tilde x}_{_{1}}} \wedge d{\tilde \psi} -\frac{1}{2} {\tilde x}_{_{2}} e^{2 y} ~ d {\tilde \psi} \wedge d {\tilde \psi}.\label{3.64}
\end{eqnarray}
The scalar curvature of the dual metric is also constant and equal to the same value from the metric of original model.
Before proceeding to investigate the conformal invariance conditions of the dual background, let us evaluate the dual dilaton.
First we find that $s\hspace{-0.6mm}\det({\tilde E}^{{ab}}) = -e^{-4y}$ and $s\hspace{-0.6mm}\det({{\tilde a}^{^a}}_{~b}({\tilde g})) =1 $,
then it follows from \eqref{3.45.2} that ${\tilde \Phi} =\phi^{^{(0)}} -2y$, and hence
we get the dilaton by remembering that $\phi^{^{(0)}}= {\cal C}_{_0} + y$ which gives the final result
\begin{eqnarray}\label{3.65}
{\tilde \Phi} = {\cal C}_{_0} - y.
\end{eqnarray}
One can check that the field equations \eqref{3.22}-\eqref{3.24} are satisfied for the metric
\eqref{3.63}, the tensor field \eqref{3.64} and
the dilaton field \eqref{3.65} together with a zero cosmological constant.
This means that the dual background is also conformally invariant up to one-loop order.
In order to get more insight of the dual metric one may use the transformation
\begin{eqnarray}
e^{y} = \frac{1}{{r}^2},~~~~~~~~~~~ {{\tilde x}_{_{1}}} = \varphi,~~~~~~~~~ {{\tilde x}_{_{2}}}= \frac{1}{l} t,   \label{3.66}
\end{eqnarray}
then, the dual background takes the  following form
\begin{eqnarray}
{\tilde ds^2}&=& \frac{l^2}{r^2}  dr^2 - \frac{r^2}{l^2} dt^2 +r^2 d\varphi^2 -\frac{2}{r^2} d {\tilde \psi } d{\tilde \chi}, \nonumber\\
{\tilde B} &=& {\tilde \psi }~  d\varphi \wedge d {\tilde \psi }- \frac{1}{2 l r^4} t~ d {\tilde \psi } \wedge d {\tilde \psi },\nonumber\\
{\tilde \Phi} &=& {\cal C}_{_0} +2 \log r. \label{3.67}
\end{eqnarray}
As can be clearly seen, the bosonic part of the metric remains unchanged under the non-Abelian T-duality.
This means that the dual model metric also describes the BTZ vacuum metric coupled to two fermionic fields.

Starting with the superdouble $\big(({\C}^3 +{\A}) , {\cal I}_{_{(2|2)}}\big)$ we thus illustrated a concrete example of the super PL T-duality and
found another non-Abelian target space dual for the BTZ vacuum metric coupled to two fermionic fields in the form
of equations \eqref{3.63}-\eqref{3.65}.

\subsection{Non-Abelian target space dual starting from the ${C}^3 \oplus {A}_{1,1}$ Lie supergroup}
\label{IV.3}
Here the Lie supergroup of target space supermanifold  ${{\M}} \approx O \times  G$
is considered to be ${C}^3 \oplus { A}_{1,1}$ whose Lie superalgebra is decomposable and it belongs to Family III.
 According to Table 4 of Appendix A, only the commutator of B-F for this
superalgebra is non-zero.
The Lie superalgebra of semie-Abelian Drinfeld superdouble
$\big({\C}^3 \oplus {\A}_{1,1}\;,\;{\cal I}_{_{(2|2)}}\big)$ is defined by
the following (anti-)commutation relations
\begin{eqnarray}\label{3.68}
{[T_{_1} , T_{_4}]} = T_{_3},~~~~~~~~~~{[T_{_1} , {\tilde T}^3]} = -{\tilde T}^4,~~~~~~~~\{T_{_4} , {\tilde T}^3\}=-{\tilde T}^1.
\end{eqnarray}
The parametrization of a general element of ${C}^3 \oplus {A}_{1,1}$ we choose as in \eqref{3.54}, giving us
\begin{eqnarray}\label{3.69}
R_{\pm}^1&=& \partial_{\pm} x_{_{1}},~~~~~~~~~~~~~~~~~~~~~~~~R_{\pm}^2 ~=~ \partial_{\pm} x_{_{2}},\nonumber\\
R_{\pm}^3&=& -\partial_{\pm} \psi- \partial_{\pm} \chi~  x_{_{1}}, ~~~~~~~~~~R_{\pm}^4 ~=~ -\partial_{\pm} \chi.
\end{eqnarray}
By a convenient choice of the spectator-dependent matrices in the form of
\begin{eqnarray}
{E_{0}}_{_{ab}}=\left( \begin{tabular}{cccc}
                 $e^y$ & 0 & 0 & 0 \\
                 0 & -$e^y$  & 0& 0 \\
                 0 & 0& $0$ & $e^y$ \\
                 0 & 0 & -$e^y$ & 0 \\
                 \end{tabular} \right),~~~~~F^{(1)}_{a i}  = F^{(2)}_{i a} = 0,~~~~F_{i j} = \frac{l^2}{4},\label{3.70}
\end{eqnarray}
and then with the help of relations \eqref{3.4} and \eqref{3.8} one can construct the
action of original $\sigma$-model on the ${C}^3 \oplus {A}_{1,1}$.
The corresponding background including the line element and zero $B$-field is given by
\begin{eqnarray}
ds^2&=& \frac{l^2}{4}  dy^2 + e^y \Big[{d x_{_{1}}}^2  - ~ {d x_{_{2}}}^2 -2  ~ d\psi d\chi\Big],\nonumber\\
B&=& 0.\label{3.71}
\end{eqnarray}
The metric is flat in the sense that both its scalar curvature and Ricci tensor vanish.
The bosonic part of the above metric (the first three terms) is similar to that of metric \eqref{3.57}, therefore,
one may use the transformation \eqref{3.60} to conclude that the metric in \eqref{3.71} is nothing but
the BTZ vacuum metric coupled to two fermionic fields. In addition, the dilaton field
that supports the background is found from \eqref{3.45.1} to be $\Phi = \phi^{^{(0)}} = {\cal C}_{_0}$.
Using these, one verifies the one-loop beta-function equations with zero cosmological constant.

Similar to the construction of dual $\sigma$-models in the previous subsections,
the Lie supergroup $G$ of dual supermanifold is here assumed to be
Abelian. With this in mind, one finds that the only non-zero component of the super Poisson structure on the dual Lie supergroup is
${\tilde \Pi}_{{{\tilde x}_{_{1}} {\tilde \chi}}} = -{\tilde \psi}$. Moreover, using the first equation of
\eqref{3.11} one gets $s\hspace{-0.6mm}\det ({{{\tilde E}^{ab}}})=-1$. We will use this in calculating the dual dilaton field in the next.
The corresponding elements to the dual model can be obtained by making use of relations \eqref{3.70}
and \eqref{3.11}. They are then read
\begin{eqnarray}
{\tilde ds^2} &=& \frac{l^2}{4}  dy^2 + e^{-y} \Big[{d {\tilde x}_{_{1}}}^2
- {d {\tilde x}_{_{2}}}^2 - 2 d {\tilde \psi} d{\tilde \chi}\Big],\nonumber\\
{\tilde B}&=& {\tilde \psi} e^{-2y}~  {d {\tilde x}_{_{1}}} \wedge d{\tilde \psi}.\label{3.72}
\end{eqnarray}
By transforming coordinate $y \rightarrow -y$ (without changing the rest of the coordinates)
it can be easily shown that the dual metric turns into the same metric as the original model
in \eqref{3.71}. Hence, it is said that the metric is self-dual.
In order to investigate the conformal invariance conditions of the dual background
one may use \eqref{3.45.2} to obtain ${\tilde \Phi}={\cal C}_{_0}$. Using these, one verifies
the field equations \eqref{3.22}-\eqref{3.24} for dual background \eqref{3.72} with  zero cosmological constant.

\subsection{Non-Abelian target space dual starting from the ${D}^{10}_{p=\pm\frac{1}{2}}$ Lie supergroup}

There is a possibility that we can also perform the dualization on the ${D}^{10}_{p=\pm\frac{1}{2}}$
Lie supergroup to construct other non-Abelian
target space dual for the BTZ vacuum metric coupling to two fermionic fields.
The corresponding Lie superalgebra to this supergroup belongs to Familiy IV of Table 5, where only the commutator of F-F among all commutators is zero.
In this subsection we deal with T-dual $\sigma$-models constructing out on the
semie-Abelian Drinfeld superdouble
$\big({\D}^{10}_{p} , {\cal I}_{_{(2|2)}}\big)$ whose Lie superalgebra is defined by the following  non-zero Lie superbrackets
\begin{eqnarray}\label{3.73}
{[T_{_1} , T_{_2}]} = T_{_2},~~~~~~~~[T_{_1} , T_{_3}]=(p+1) T_{_3},~~~~~[T_{_1} , T_{_4}]=p T_{_4},~~~~~~[T_{_2} , T_{_4}]=T_{_3},~~~~~~~~\nonumber\\
{[T_{_1} , {\tilde T}^2]}=-{\tilde T}^2,~~~~[T_{_1} , {\tilde T}^3]=-(p+1){\tilde T}^3,~~~[T_{_1} , {\tilde T}^4]=-p{\tilde T}^4,~~~[T_{_2} , {\tilde T}^2]={\tilde T}^1,~~~~~~~\nonumber\\
{[T_{_2} , {\tilde T}^3]} = -{\tilde T}^4,~~~\{T_{_3} , {\tilde T}^3\}=-(p+1){\tilde T}^1,~~\{T_{_4} , {\tilde T}^3\}=-{\tilde T}^2,~~~
\{T_{_4} , {\tilde T}^4\}=-p{\tilde T}^1.~~~
\end{eqnarray}
To obtain the right-invariant super one-forms we parameterize an element of the ${D}^{10}_{p}$ as in \eqref{3.54}.
We then find
\begin{eqnarray}\label{3.74}
R_{\pm}^1&=& \partial_{\pm} x_{_{1}},~~~~~~~~~~~~~~~~~~~~~~~~~~~~~~~~~~~~~~~~~~~~~~~R_{\pm}^2 ~=~ e^{x_{_{1}}}~\partial_{\pm} x_{_{2}},\nonumber\\
R_{\pm}^3&=&  \partial_{\pm} x_{_{1}} (p+1) \psi -\partial_{\pm} \psi- \partial_{\pm} \chi~  x_{_{2}} e^{(p+1) x_{_{1}}},
 ~~~R_{\pm}^4 ~=~ -\partial_{\pm} \chi~e^{p x_{_{1}}}.
\end{eqnarray}
In this way, the BTZ vacuum metric coupled to two fermionic fields may be yielded from
the original $\sigma$-model on the superdouble $\big({\D}^{10}_{p} , {\cal I}_{_{(2|2)}}\big)$ if one considers
the spectator-dependent matrices as in \eqref{3.42} by setting $\alpha_{_0} =0$.
Using these and employing \eqref{3.4} we then get
\begin{eqnarray}
ds^2&=& l^2  dy^2 + e^{{x_{_{1}}} -2y} ~ d x_{_{1}}  d x_{_{2}} +2(p+1) h(y)   \psi   e^{p{x_{_{1}}}}~ d x_{_{1}}  d\chi
- 2 h(y) e^{p x_{_{1}}} ~ d\psi d\chi,\nonumber\\
B&=& 0.\label{3.75}
\end{eqnarray}
Notice that the bosonic part of the metric is also similar to that of metric \eqref{3.43}.
Here the field strength is absent, and hence with a constant dilton field one guarantees the conformal invariance of the background \eqref{3.75}
provided that we have $h(y) = e^{-2y}$ as well as $p=\pm 1/2$.
It follows from this result that the metric is flat in the sense that its scalar
curvature vanishes.
In the following we will focus on the $p=- 1/2$ case of the superdouble only.

Before proceeding to construct the dual background, let us calculate the dual couplings.
Utilizing relation \eqref{3.10} for untilded quantities and also employing \eqref{3.73} we get
\begin{eqnarray}
{\tilde \Pi}_{ab}=\left( \begin{tabular}{cccc}
                 0 &-${\tilde x}_{_2}$ & -$\frac{1}{2} {\tilde \psi}$ & $\frac{1}{2} {\tilde \chi}$ \\
                 ${\tilde x}_{_2}$ & $0$  & 0& -${\tilde \psi}$ \\
                 $\frac{1}{2} {\tilde \psi}$ & 0& $0$ & 0 \\
                 -$\frac{1}{2} {\tilde \chi}$ & ${\tilde \psi}$ & 0 & 0\\
                 \end{tabular} \right),\label{3.76}
\end{eqnarray}
then, the dual coupling matrices are worked out
\begin{eqnarray}
{\tilde E}^{{ab}} =\left( \begin{tabular}{cccc}
                 0 & $\frac{1}{\Delta_{_+}}$ & -$ \frac{{\tilde \psi} e^{2y}}{\Delta_{_+}}$ & 0 \\
                 -$\frac{1}{\Delta_{_-}}$ & $\frac{{\tilde \psi} {\tilde \chi} e^{2y}}{2 \Delta}$  & -$ \frac{{\tilde \chi} e^{2y}}{ 2 \Delta_{_-}}$ & -$ \frac{{\tilde \psi} e^{2y}}{2 \Delta_{_-}}$ \\
                 -$ \frac{{\tilde \psi} e^{2y}}{\Delta_{_-}}$  & -$ \frac{{\tilde \chi} e^{2y}}{2 \Delta_{_+}}$ & -$\frac{{{\tilde x}_{_2}} {\tilde \psi} {\tilde \chi} e^{4y}}{\Delta}$ & $e^{2 y}$ \\
                 0 & -$ \frac{{\tilde \psi} e^{2y}}{2 \Delta_{_+}}$ & -$e^{2 y}$ & 0 \\
                 \end{tabular} \right),~~~{\tilde \phi}^{\hspace{0mm}{(1)^{ a}}}_{~~~j} = {\tilde \phi}^{\hspace{0mm}{(2)^{ b}}}_{i} = 0,~~~~~~
{\tilde \phi}_{_{ij}} =l^2,~~\label{3.77}
\end{eqnarray}
where $\Delta_{_\pm} ={{\tilde x}_{_2}} \pm \frac{1}{2} e^{-2y}$ and $\Delta =\Delta_{_+} \Delta_{_-}$.
We need to have the superdeterminant of matrix ${\tilde E}^{{ab}}$. One immediately finds that $s\hspace{-0.6mm}\det({\tilde E}^{{ab}}) = e^{-4y}/\Delta$.
Finally, the corresponding  elements to the dual model can be obtained by making use of relations \eqref{3.76} and \eqref{3.77}
together with \eqref{3.7}, giving us
\begin{eqnarray}
{\tilde ds^2}&=& {l^2}  dy^2 + \frac{1}{\Delta}\Big[ \frac{1}{2} e^{2y} {\tilde \psi} {\tilde \chi}~ {d {\tilde x}_{_2}}^2 -e^{-2y}~  d{\tilde x}_{_1} d {\tilde x}_{_2} + 2 {\tilde x}_{_2} {\tilde \psi} e^{2y} ~ d {\tilde x}_{_1} d {\tilde \psi} \nonumber\\
&&~~~~~~~~~~~~~~~~~~~~~~~~~~~~+  e^{2y} {\tilde x}_{_2}  {\tilde \chi} ~d {\tilde x}_{_2} d {\tilde \psi}
+  e^{2y} {\tilde x}_{_2}  {\tilde \psi}~ d {\tilde x}_{_2} d {\tilde \chi}\Big] - 2e^{2y} ~d {\tilde \psi}  d {\tilde \chi},\label{3.78}\\
{\tilde B}&=& \frac{1}{\Delta}\Big[{\tilde x}_{_2}  ~ d{\tilde x}_{_1}  \wedge d {\tilde x}_{_2} -\frac{1}{2} {\tilde \psi} ~d {\tilde x}_{_1}\wedge  d {\tilde \psi}+\frac{1}{4} {\tilde \chi} ~d {\tilde x}_{_2}\wedge  d {\tilde \psi}+\frac{1}{4} {\tilde \psi} ~d {\tilde x}_{_2}\wedge  d {\tilde \chi}
\nonumber\\
&&~~~~~~~~~~~~~~~~~~~~~~~~~
~~~~~~~~~~~~~~~~~~~~~~~~~~~+\frac{1}{2} {\tilde x}_{_2} e^{4 y}  {\tilde \psi} {\tilde \chi}~ d {\tilde \psi} \wedge  d {\tilde \psi}\Big].\label{3.79}
\end{eqnarray}
The scalar curvature of the dual metric is also zero.
It seems that under the non-Abelian T-duality the scalar curvature
has been restored from the dual model to the original one.
From equation \eqref{3.45.2} and also from the superdeterminant of matrix ${\tilde E}^{{ab}}$ that found above,
one gets the dual dilaton field, giving ${\tilde \Phi} = {\cal C}_{_0}-2 y -\frac{1}{2}\log |\Delta|$.
Unfortunately, the field equations \eqref{3.22}-\eqref{3.24} do not satisfy with the metric
\eqref{3.78}, $B$-field \eqref{3.79} and the above dilaton.
In fact, unlike the original model, the dual model cannot be conformally invariant up to one-loop order.

Let us discuss below the reason for not preserving conformal invariance under non-Abelian T-duality for this example.
In Ref. \cite{Sfetsos1}, Sfetsos proved the classical canonical equivalence to the $\sigma$-models related by PL T-duality.
The canonical transformations are essentially classical and
their equivalence can hold in some special cases but it fails in most cases.
However, checking the equivalence by studying conformal
invariance up to one-loop order is important.
On the other hand, it has been shown that \cite{N.Mohammedi} a sufficient condition for the invariance of
the reduced string effective action under PL T-duality is the vanishing of the traces of the structure constants
of each Lie algebra constituting the Drinfeld double.
Notice that the field equations \eqref{3.22}-\eqref{3.24} can be interpreted as field equations for ${G}_{_{MN}}$,
${B}_{_{MN}}$ and $\Phi$ of the low energy string effective action.
The string tree level effective action on a $d$-dimensional supermanifold $\M$ for these background fields is given by \cite{ER5}
\begin{eqnarray}\label{3.80}
S_{eff}&=&\int d^{^d}X\; \sqrt{-G}e^{-2\Phi}  {\cal L}_{_{eff}}\nonumber\\
&=&\int d^{^d}X\; \sqrt{-G}e^{-2\Phi} \Big[{\cal R}+4{\overrightarrow{\nabla}_{_M}} {\Phi} \overrightarrow{\nabla}^{^M} {\Phi}
+\frac{1}{12} H_{_{MNP}}H^{^{PNM}}-4\Lambda\Big],
\end{eqnarray}
where $G$ stands for the superdeterminant of $_{_{M}}{G}_{_{N}}$.
In the case of the above example, one first finds the supertrace (str) of adjoint representations ${{({\cal X}_b)}_c}^{a}=-{f^a}_{bc}$ of  ${\D}^{10}_{p=-\frac{1}{2}}$, giving us,
str${({\cal X}_1)}=-1$, str${({\cal X}_i)}= 0,~ i=2,3,4$. Then, by calculating the effective Lagrangians corresponding to both original and dual models,
we find that the dual Lagrangian is not invariant under super PL T-duality transformation and therefore
both Lagrangians are not equal. Obviously, this anomaly is due to the non-vanishing
traces of the structure constants of the superdouble $\big({\D}^{10}_{p} , {\cal I}_{_{(2|2)}}\big)$.
Moreover, in this example one can show that the integration weights $\sqrt{-G} e^{-2 \Phi}$
and $\sqrt{- \tilde G} e^{-2 \tilde \Phi}$ are not equal.
 The reason behind this may be due to the particularity of our model.
There is a possibility of absorbing the anomalous terms into
dilaton shift which is the same as a diffeomorphism transformation. We note that the equations
\eqref{3.45.1} and \eqref{3.45.2} are only transformations which lead to a proportionality
between the integration weights $\sqrt{-G} e^{-2\Phi}$ and $\sqrt{- \tilde G} e^{-2 \tilde \Phi}$ \cite{N.Mohammedi}.

\subsection{Non-Abelian target space dual starting from the $(2{ A}_{1,1}+2{ A})^0$ Lie supergroup}

As the last example of this section we shall consider the $(2{\A}_{1,1}+2{ \A})^0$ Lie superalgebra of Familiy V which is defined by one
non-zero superbracket only as presented in Table 6.
First of all, let us define the Lie superalgebra of semie-Abelian Drinfeld superdouble
$\big((2{\A}_{1,1}+2{ \A})^0 , {\cal I}_{_{(2|2)}}\big)$. The generators of $(2{\A}_{1,1}+2{ \A})^0$ and ${\cal I}_{_{(2|2)}}$ are denoted,
respectively, $T_a$ and ${\tilde T}^a$, $a=1, . . . , 4$ and
satisfy the (anti-)commutation relations
\begin{eqnarray}\label{3.81}
\{T_{_3} , T_{_3}\} = T_{_1},~~~~~~~~~~~[T_{_3} , {\tilde T}^1]=-{\tilde T}^3.
\end{eqnarray}
Choosing a parametrization as in \eqref{3.40} for the elements of the $(2{ A}_{1,1}+2{ A})^0$ Lie supergroup leads to the following right-invariant super one-forms
\begin{eqnarray}\label{3.82}
R_{\pm}^1= \partial_{\pm} x_{_{1}} + \partial_{\pm} \psi ~\frac{\psi}{2},~~~~~~R_{\pm}^2 = \partial_{\pm} x_{_{2}},~~~~~~
R_{\pm}^3 = -\partial_{\pm} \psi,~~~~~R_{\pm}^4 = -\partial_{\pm} \chi.
\end{eqnarray}
Here we choose the spectator-dependent matrices as in \eqref{3.56} by setting $a_{_1} (y)= -a_{_2} (y)= e^y,~a_{_3} (y)= e^{-y}$ and $b(y) =0$.
Thus, the corresponding elements to the original $\sigma$-model including the line element and $B$-field may be expressed as
\begin{eqnarray}
ds^2&=& \frac{l^2}{4}  dy^2 + e^y \Big[{d x_{_{1}}}^2  - ~ {d x_{_{2}}}^2 -\psi ~ {d x_{_{1}}} d\psi\Big] -2  e^{-y}~ d\psi d\chi,\nonumber\\
B&=& 0.\label{3.83}
\end{eqnarray}
The metric is flat in the sense that its scalar curvature is ${\cal R}=-16/l^2$.
Moreover, one can easily show that the background \eqref{3.83} is conformally
invariant up to one-loop order with zero cosmological constant and  dilaton $\Phi ={\cal C}_{_0}+y$.
From comparison with relation \eqref{3.45.1}, we conclude that $\phi^{^{(0)}}={\cal C}_{_0}+y$.
Again one can use coordinate transformation \eqref{3.60} to rewrite the metric in \eqref{3.83} in the form of
the BTZ vacuum metric coupling to two fermionic fields.

As already mentioned, the super Poisson structure on the dual Lie supergroup can be obtained by making use of the dual version of equation \eqref{3.10}.
Using (anti-)commutation relations of the superdouble in \eqref{3.81}, one finds that
the only non-zero component of the dual super Poisson structure is
${\tilde \Pi}_{{{\tilde \psi} {\tilde \psi}}} = -{{\tilde x}_{_1}}$. Finally, we can obtain the
dual background corresponding to \eqref{3.83}, in such a way that
the line element and the tensor
field ${\tilde B}$ take the following forms
\begin{eqnarray}
{\tilde ds^2}&=& \frac{l^2}{4}  dy^2 + e^{-y}  \big({d {\tilde x}_{_{1}}}^2  - {d {\tilde x}_{_{2}}}^2\big) - 2 e^{y} ~ d {\tilde \psi} d{\tilde \chi},\label{3.84}\\
{\tilde B}&=& -\frac{1}{2} {\tilde x}_{_{1}} e^{2 y} ~ d {\tilde \chi} \wedge d {\tilde \chi}.\label{3.85}
\end{eqnarray}
The scalar curvature of the metric is ${\tilde {\cal R}}=-16/l^2$ just as the original model; moreover, one finds that the only
non-zero components of field strength are ${\tilde H}_{_{y {\tilde \chi} {\tilde \chi}}}=2 {\tilde x}_{_{1}} e^{2 y},~{\tilde H}_{_{{\tilde x}_{_{1}} {\tilde \chi} {\tilde \chi}}}=e^{2 y}$. Also, the dual dilaton is obtained from \eqref{3.45.2} to be ${\tilde \Phi} ={\cal C}_{_0} - y$. Using these, one verifies the
field equations \eqref{3.22}-\eqref{3.24} with a zero cosmological constant.

For the sake of clarity the results obtained in this section are summarized in Table
1; we display non-Abelian target space duals of the  BTZ vacuum metric coupled to two fermionic fields
together with background of original models and their transformed.

\section{Super PL T-plurality of BTZ vacuum metric coupled to two fermionic fields}
\label{Sec.V}

In the preceding section, we studied the non-Abelian T-dualization of the BTZ vacuum metric coupling to two fermionic fields,
in such a way that we worked with semi-Abelian superdoubles which were non-isomorphic as Lie superalgebras in each of the models.
In this section we first show that for the ${\C}^3 \oplus {\A}_{1,1}$ Lie superalgebra
there are several Manin supertriples and the possibility of embedding these Manin supertriples into the corresponding
Drinfeld superdoubles. We also obtained a non-Abelian dual for the BTZ vacuum metric coupled to two fermionic fields
by constructing mutually T-dual $\sigma$-models on the Manin supertriple $({\C}^3 \oplus {\A}_{1,1} , {\cal I}_{_{(2|2)}})$.
Here we shall obtain the conformal duality chain of cosmological string backgrounds
with non-vanishing torsion on the other Manin supertriples in corresponding Drinfeld superdoubles.
\begin{center}
		\small{{{\bf Table 1.}~ Non-Abelian target space duals of the  BTZ vacuum metric \\coupled to two fermionic fields$~~~~~~~~~~~~~~~~~~~~~$}}
		{\scriptsize
			\renewcommand{\arraystretch}{1.5}{
\begin{tabular}{| l|l|l| l|} \hline \hline
Superdouble & Original background & Transformed  background & Dual background \\
 ${\D}=({\G} , {\tilde {\G}})$& & of original model &  \\ \hline
 & $ds^2= l^2  dy^2 + e^{{x_{_{1}}} -2y}\big[d x_{_{1}}  d x_{_{2}}$&  $ds^2 = \frac{l^2}{r^2}  dr^2 - \frac{r^2}{l^2} dt^2 +r^2 d\varphi^2$ & ${\tilde ds^2} = {l^2}  dy^2 + \frac{1}{\Delta}\Big[-e^{-2y}  d{\tilde x}_{_1} d {\tilde x}_{_2}$  \\
 &  $ - \psi  ~ d x_{_{1}}  d\chi +d\psi d\chi\big],$  &$ + \frac{r^2}{l^2}\Big[ -\frac{\psi}{l}  dt d\chi+\psi d\varphi d\chi$ &  $+ \frac{\alpha_{_0}}{\Delta}  (1-\frac{1}{4} e^{-4 y}) {\tilde \psi} d {\tilde x}_{_2} d {\tilde \psi} $\\
 $\big(({\C}^1_1 +{\A}) , {\cal I}_{_{(2|2)}}\big)$ &  &$+d\psi d\chi \Big],$ & $+ \frac{2}{\Delta} e^{-2y} \big({\tilde x}_{_2}  {\tilde \psi} d {\tilde x}_{_2} d {\tilde \chi}- d {\tilde \psi}  d {\tilde \chi}\big)\Big],$ \\	
  & $B=-\frac{\alpha_{_0}}{2} (1-\frac{1}{4} e^{-4y}) d\chi \wedge d\chi,$ &$B= -\frac{\alpha_{_0}}{ 2}(1-\frac{r^4}{4l^4})d\chi \wedge d\chi,$ & ${\tilde B}= \frac{1}{\Delta}\Big[{\tilde x}_{_2}  d{\tilde x}_{_1}  \wedge d {\tilde x}_{_2} -{\tilde x}_{_2} d{\tilde \psi} \wedge  d {\tilde \chi}$\\
 &  & & $-\frac{\alpha_{_0} (4-{e^{-4 y}})}{8}  d{\tilde \psi} \wedge d{\tilde \psi} +d {\tilde x}_{_2}  \wedge  d\eta\Big], $\\	
  & $\Phi = {\cal C}_{_0}$ &$\Phi = {\cal C}_{_0}$ & ${\tilde \Phi}= {\cal C}_{_0}$ \\	\hline

 & $ds^2= \frac{l^2}{4}  dy^2 + e^{y} \big({d x_{_{1}}}^2 - {d x_{_{2}}}^2 $&  $ds^2 = \frac{l^2}{r^2}  dr^2 - \frac{r^2}{l^2} dt^2+r^2 d\varphi^2 $ & ${\tilde ds^2} = \frac{l^2}{4}  dy^2 + e^{-y} ~ {d {\tilde x}_{_{1}}}^2$  \\
 &  $ +  \chi d x_{_{2}} d \chi\big)-2 e^{-y} ~ d\psi d\chi,  $  &$ + \frac{r^2}{l} \chi dt d\chi -\frac{2}{r^2} d\psi d\chi,$ &  $-e^{-y} ~ {d {\tilde x}_{_{2}}}^2 - 2 e^{y} ~ d {\tilde \psi} d{\tilde \chi},$\\
  $\big(({\C}^3 +{\A}) , {\cal I}_{_{(2|2)}}\big)$&  & & ${\tilde B}= {\tilde \psi} ~  {d {\tilde x}_{_{1}}} \wedge d{\tilde \psi}$ \\	
  & $B=-\frac{1}{2} b(y)~ d\chi \wedge d\chi,$ &$B= -\frac{\beta_{_0}}{2} (1-\frac{1}{2r^4}) d\chi \wedge d\chi,$ & $-\frac{1}{2} {\tilde x}_{_{2}} e^{2 y} ~ d {\tilde \psi} \wedge d {\tilde \psi},$\\	
  & $\Phi = {\cal C}_{_0} + y$ &$\Phi = {\cal C}_{_0} +2 \log r$ & ${\tilde \Phi}= {\cal C}_{_0} - y$ \\	\hline

 & $ds^2= \frac{l^2}{4}  dy^2 + e^{y} \big({d x_{_{1}}}^2 - {d x_{_{2}}}^2 $&  $ds^2 = \frac{l^2}{r^2}  dr^2 - \frac{r^2}{l^2} dt^2+r^2 d\varphi^2 $ & ${\tilde ds^2} = \frac{l^2}{4}  dy^2 + e^{-y} \big[{d {\tilde x}_{_{1}}}^2$  \\
 &  $ -2  d\psi d\chi\big),  $  &$ -{2}{r^2} d\psi d\chi,$ &  $- {d {\tilde x}_{_{2}}}^2 - 2  d {\tilde \psi} d{\tilde \chi}\big],$\\
$\big({\C}^3 \oplus {\A}_{1,1} , {\cal I}_{_{(2|2)}}\big)$  & $B=0,$ & $B=0,$ & ${\tilde B}= {\tilde \psi}  e^{-2 y}~  {d {\tilde x}_{_{1}}} \wedge d{\tilde \psi},$ \\	
  & $\Phi = {\cal C}_{_0}$ &$\Phi = {\cal C}_{_0}$ & ${\tilde \Phi}= {\cal C}_{_0}$ \\	\hline

& $ds^2= l^2  dy^2 + e^{{x_{_{1}}} -2y} d x_{_{1}}  d x_{_{2}}$&  $ds^2 = \frac{l^2}{r^2}  dr^2 - \frac{r^2}{l^2} dt^2 +r^2 d\varphi^2$ & ${\tilde ds^2} = {l^2}  dy^2   +\frac{1}{\Delta}\Big[\frac{1}{2} e^{2y} {\tilde \psi} {\tilde \chi}{d {\tilde x}_{_2}}^2  $  \\
 &  $ +e^{-(\frac{{x_{_{1}}}}{2} +2y)} \big(\psi  d x_{_{1}}  d\chi$  &$ + {r^2}\Big(l(t-l \varphi)\Big)^{-\frac{3}{2}} \Big[ dt \psi d\chi$ &  $-e^{-2y}  d{\tilde x}_{_1} d {\tilde x}_{_2}+ 2 {\tilde x}_{_2} {\tilde \psi} e^{2y} ~ d {\tilde x}_{_1} d {\tilde \psi}$\\
   &  $ -2 d\psi d\chi\big),$&$-l d\varphi \psi d\chi\Big]$ & $+  e^{2y} {\tilde x}_{_2}  {\tilde \chi} d {\tilde x}_{_2} d {\tilde \psi}+  e^{2y} {\tilde x}_{_2}  {\tilde \psi} d {\tilde x}_{_2} d {\tilde \chi}\Big]$ \\
 $\big({\D}^{10}_{p=-\frac{1}{2}} , {\cal I}_{_{(2|2)}}\big)$ &  &$-2 \frac{r^2}{l} \big[l(t-l \varphi)\big]^{-\frac{1}{2}} d\psi d\chi,$ & $- 2e^{2y} d {\tilde \psi}  d {\tilde \chi},$ \\	
  & $B=0,$ &$B= 0,$ & ${\tilde B}= \frac{1}{\Delta}\Big[{\tilde x}_{_2} d{\tilde x}_{_1}  \wedge d {\tilde x}_{_2}-\frac{{\tilde \psi}}{2}  d {\tilde x}_{_1}\wedge  d {\tilde \psi}$\\
 &  & & $+\frac{1}{4} {\tilde \chi} ~d {\tilde x}_{_2}\wedge  d {\tilde \psi}+\frac{1}{4} {\tilde \psi} ~d {\tilde x}_{_2}\wedge  d {\tilde \chi} $\\
 	 &  & & $+\frac{1}{2} {\tilde x}_{_2} e^{4 y}  {\tilde \psi} {\tilde \chi}~ d {\tilde \psi} \wedge  d {\tilde \psi}\Big],$ \\
  & $\Phi = {\cal C}_{_0}$ &$\Phi = {\cal C}_{_0}$ & ${\tilde \Phi}= {\cal C}_{_0}-2y -\frac{1}{2} \log|\Delta|$ \\	\hline

 & $ds^2= \frac{l^2}{4}  dy^2 + e^{y} \big({d x_{_{1}}}^2 - {d x_{_{2}}}^2 $&  $ds^2 = \frac{l^2}{r^2}  dr^2 - \frac{r^2}{l^2} dt^2+r^2 d\varphi^2 $ & ${\tilde ds^2} = \frac{l^2}{4}  dy^2 + e^{-y} ~ {d {\tilde x}_{_{1}}}^2$  \\
 &  $ -  \psi d x_{_{1}} d \psi\big)-2 e^{-y} ~ d\psi d\chi,  $  &$ -{r^2} \psi d\varphi d\psi -\frac{2}{r^2} d\psi d\chi,$ &  $-e^{-y} ~ {d {\tilde x}_{_{2}}}^2 - 2 e^{y} ~ d {\tilde \psi} d{\tilde \chi},$\\
  $\big((2{\A}_{1,1}+2{ \A})^0 $&  & &  \\	
  $ , {\cal I}_{_{(2|2)}}\big)$& $B=0$ &$B= 0$ & ${\tilde B}=-\frac{1}{2} {\tilde x}_{_{1}} e^{2 y} ~ d {\tilde \chi} \wedge d {\tilde \chi},$\\	
  & $\Phi = {\cal C}_{_0} + y$ &$\Phi = {\cal C}_{_0} +2 \log r$ & ${\tilde \Phi}= {\cal C}_{_0} - y$ \\\hline 	\hline
		\end{tabular}}}
\end{center}
\vspace{-1mm}
{\scriptsize~~~~~~~~~~~$b(y)={\beta_{_0}} (1-\frac{1}{2} e^{-2y})$,~~~$\Delta ={{\tilde x}_{_2}}^2 -\frac{1}{4} e^{-4y} $,~~~$d \eta =\frac{\alpha_{_0}}{\Delta}  (1-\frac{1}{4} e^{-4 y}) {\tilde x}_{_2} {\tilde \psi}   d {\tilde \psi}
+\frac{1}{\Delta} ({{\tilde x}_{_2}}^2 +\frac{1}{4} e^{-4 y}) {\tilde \psi} d {\tilde \chi}$.}\\



\subsection{A brief review of super PL T-plurality}
\label{V.1}

To set our notation let us briefly review the construction of the super PL T-plural $\sigma$-models on Lie
supergroups \cite{Eghbali3}.
In what follows we shall consider untransformed $\sigma$-model on $G$ of the superdouble $D=(G , {\tilde G})$ in the following form
\begin{eqnarray}\label{5.1}
S = \frac{1}{2} \int_{_{\Sigma}} d\sigma^{+}  d\sigma^{-} \Big[(-1)^a  R_{+}^a(g)~{{E}_{_{ab}}}(g, y^i)~
 R_{-}^b(g) + \partial_{+} y^{i} F_{_{ij}}(y^i) \partial_{-} y^{j} \Big].
\end{eqnarray}
We start with the assumption that there are several decompositions of a Drinfeld superdouble.
Let $\mathbb{X}_{_A}=\{T_{_a} , {\tilde T}^{^b}\},~ a, b=1,...,dim \hspace{0.4mm}G$ be generators of Lie sub-superalgebras $\G$ and $\tilde {\G}$  of a Drinfeld superdouble $\D$
associated with the  $\sigma$-model \eqref{5.1}, and consider $\mathbb{X}'_{_A}=\{U_{_a} , {\tilde U}^{^b}\}$ as generators
of other Manin supertriple $({\G}_{_U} , {\tilde {\G}}_{_U})$ in the same Drinfeld superdouble ($\D \cong {\D}_{_U}$). Both
$\D$ and  ${\D}_{_U}$ satisfy equations \eqref{3.1} and \eqref{3.2}.
It is said that $\D$ and ${\D}_{_U}$ are isomorphic to each other iff there is an invertible supermatrix ${C_{_{A}}}^{^B}$ such that the linear map given by
$\mathbb{X}_{_A} = (-1)^{^B}~ {C_{_{A}}}^{^B}~ \mathbb{X}'_{_B}$ transforms the Lie multiplication of $\D$ into that of ${\D}_{_U}$ and preserves
the canonical form of the bilinear form $<. , .>$.
One may define the isomorphism  transformation between the bases of $\D$ and  ${\D}_{_U}$ in explicit components form as follows \cite{Eghbali3}:
\begin{eqnarray}
T_{_a} &=& (-1)^{^c}~ {F_{_a}}^{^c}~U_{_c} + G_{_{ac}}~{\tilde U}^{^c}, \nonumber\\
{\tilde T}^{^b} &=& (-1)^{^c}~ H^{^{bc}}~U_{_c} + {K^{^b}}_{_{c}}~{\tilde U}^{^c}.\label{5.2}
\end{eqnarray}
Now we can define a $\sigma$-model over the Lie supergroup associated with the generators $U_{_a}$.
In fact, the transformed $\sigma$-model is given by the same form as \eqref{5.1} but with
$E(g , y^i)$ replaced by ${{E}_{_{ab}}}(g_{_U}, y^i)$. It is then read
\begin{eqnarray}\label{5.3}
S_{_U} = \frac{1}{2} \int_{_{\Sigma}} d\sigma^{+}  d\sigma^{-} \Big[(-1)^a  R_{+}^a(g_{_U})~{{E}_{_{ab}}}(g_{_U}, y^i)~
 R_{-}^b(g_{_U}) + \partial_{+} y^{i} F_{_{ij}}(y^i) \partial_{-} y^{j} \Big],
\end{eqnarray}
where
\begin{eqnarray}\label{5.4}
E(g_{_{_U}})=\big(N M^{^{-1}} + \Pi(g_{_{_U}})\big)^{^{-1}},
\end{eqnarray}
in which the matrices $M, N$ and $\Pi(g_{_{_U}})$ in the form of their components may be expressed as
\begin{eqnarray}
M_{_{ab}}&=& (-1)^{^c}~ {{(K^{st})}_{_{a}}}^{{c}}~ {E_{_0}}_{_{cb}} - {(G^{st})}_{_{ab}},\label{5.5}\\
{N^{^a}}_{_{b}}~&=& {(F^{st})^{a}}_{_{b}}- (-1)^{^c}~ {(H^{st})}^{^{ac}}~ {E_{_0}}_{_{cb}},\label{5.6}\\
\Pi^{^{ab}}(g_{_{_U}})&=& (-1)^{^c}~ b^{^{ac}}(g_{_{_U}}) ~ {{(a^{-1})}_{_c}}^{^b} (g_{_{_U}}).\label{5.7}
\end{eqnarray}
Notice that the sub-matrices $a(g_{_{_U}})$ and $b(g_{_{_U}})$ is defined as in \eqref{3.10}.

In Ref. \cite{vonUnge},  von Unge showed that the plurality transformation must be supplemented by a
correction that comes from integrating out the fields on the dual group $\tilde G$
in path-integral formulation.
In some cases it can be absorbed at the one-loop level into the transformation of the dilaton field.
Following \cite{vonUnge}, the transformation of dilaton on Lie supergroup can be written as follows:
\begin{eqnarray}\label{5.8}
\Phi_{_{_U}} = \phi^{^{(0)}} +\frac{1}{2}\log \big|s\hspace{-0.6mm}\det\big({_{_a}}{E_{_b}}(g_{_{_U}})\big) \big|-
\frac{1}{2}\log \big|s\hspace{-0.6mm}\det\big({_{_a}}{M_{_b}}\big) \big|+
\frac{1}{2}\log \Big|s\hspace{-0.6mm}\det\Big({_{_a}}{(a^{-1})^{^b}} {\tiny (g_{_{_U}})}\Big)\Big|,~~~
\end{eqnarray}
in which $\phi^{^{(0)}}$ is the dilaton making the original $\sigma$-model conformal and may depend on the coordinates of $G_{_U}$.
Since the super PL T-duality is a special case of the super plurality,  one may consider
$F=K={\bf 1}, G=H=0$ in \eqref{5.2}
to obtain ${N^{^a}}_{_{b}} = {\delta^{^a}}_{_{b}}, M_{_{ab}} = {E_{_{0}}}_{ab}$.
Plugging this into \eqref{5.8} we arrive at \eqref{3.45.1}.
In the following we employ the above formulation to study super PL T-plurality of the BTZ vacuum metric coupled to two fermionic fields.

\subsection{Super PL T-plurality with respect to the ${C}^3 \oplus {A}_{1,1}$ Lie supergroup}

So far, there has been no classification of the Lie superbialgebras structures on the ${\C}^3 \oplus {\A}_{1,1}$.
We have presented the general form of the structure constants related to the dual Lie superalgebras to
${\C}^3 \oplus {\A}_{1,1}$ in Appendix B.
Here we shall present only those occurring in
this work, i.e., only first class including the following four isomorphic Manin supertriples:
\begin{eqnarray}\label{5.9}
&&~~\big({\C}^3 \oplus {\A}_{1,1} , {\cal I}_{_{(2|2)}}\big) \cong  \big({\C}^3 \oplus {\A}_{1,1} , {\C}^3 \oplus {\A}_{1,1}.i\big)
\cong \big({\C}^3 \oplus {\A}_{1,1} , {(2{\A}_{1,1}+2{ \A})^0}.i\big) \nonumber\\
&&~~~~~~~~~~~~~~~~~~~~~~~~~\cong  \big({\C}^3 \oplus {\A}_{1,1} , (({\A}_{1,1}+2{ \A})^2 \oplus {\A}_{1,1}).i\big).
\end{eqnarray}
In the following, we will present the isomorphism transformation between these superdoubles.
Referring to solutions of the super Jacobi and mixed super Jacobi
identities listed in Appendix B, if one sets $\alpha=\gamma=0, \beta \neq 0$ in case (1) or case (2) of the solutions, then it is
isomorphic to the ${\C}^3 \oplus {\A}_{1,1}$. We denote this isomorphic Lie superalgebra by ${\C}^3 \oplus {\A}_{1,1}.i$.
To obtain the dual Lie superalgebra ${(2{\A}_{1,1}+2{ \A})^0}.i$ of the Manin supertriples
$\big({\C}^3 \oplus {\A}_{1,1} , {(2{\A}_{1,1}+2{ \A})^0}.i\big)$ one may set $\alpha=\beta=0, \gamma \neq 0$ in case (2)
or $\alpha=\gamma=0, \beta \neq 0$ in case (3); moreover, this superalgebra can be obtained from case (4) by putting
$\beta=\gamma=\lambda=0, \alpha \neq 0$ or $\alpha=\beta=0, \gamma \neq 0$ in case (5).
Finally, the $(({\A}_{1,1}+2{ \A})^2 \oplus {\A}_{1,1}).i$
is obtained from case (3) by putting $\alpha =0, \beta\neq 0, \gamma \neq 0$; moreover, it can be obtained from case (4)
by putting $\alpha = \beta= \lambda=0,  \gamma \neq 0$.
\\\\
\smallskip
$\bullet$~$\D={\bf \big({\C}^3 \oplus {\A}_{1,1} , {\cal I}_{_{(2|2)}}\big)}$:
\smallskip
\smallskip
In order to obtain a possible conformal duality chain on the isomorphic Drinfeld superdoubles of \eqref{5.9}
we begin with $\D=({\C}^3 \oplus {\A}_{1,1} , {\cal I}_{_{(2|2)}})$
whose Lie superalgebra has been given in \eqref{3.68}.
The corresponding right-invariant super one-forms and
constant matrix $E_{_0}$ illustrated in subsection \ref{IV.3}. Accordingly,
the action of  $\sigma$-model on the ${C}^3 \oplus {A}_{1,1}$ in the form \eqref{5.1} is given by
\begin{eqnarray}\label{5.10}
S = \frac{1}{2} \int_{_{\Sigma}} d\sigma^{+}  d\sigma^{-} \Big[\frac{l^2}{4} ~\partial_+y \partial_-y +e^y~\big(\partial_+ {x_{_{1}}}
\partial_- {x_{_{1}}} - \partial_+ {x_{_{2}}}
\partial_- {x_{_{2}}} +\partial_+ {\psi} \partial_- {\chi}- \partial_+ {\chi} \partial_- {\psi}\big)\Big].~~~~~~~
\end{eqnarray}
Notice that under the transformation \eqref{3.60}, the background of action (equation \eqref{3.71}) describes
the BTZ vacuum metric coupling to two fermionic fields as was mentioned in subsection \ref{IV.3}.
In addition, it was shown there that the dilton field guaranteeing the conformal invariance of the model is $\Phi = \phi^{^{(0)}} = {\cal C}_{_0}$.
\\\\
\smallskip
$\bullet$~${\D}_{_U}={\bf \big({\C}^3 \oplus {\A}_{1,1} , {\C}^3 \oplus {\A}_{1,1}.i\big)}$:
\smallskip
\smallskip
\\
The Lie superalgebra of the superdouble ${\D}_{_U}=\big({\C}^3 \oplus {\A}_{1,1} , {\C}^3 \oplus {\A}_{1,1}.i\big)$
obeys the following set of non-trivial (anti-)commutation relations:
\begin{eqnarray}
[U_{_1} , U_{_4}] &=&U_{_3},~~~~~~~~~[{\tilde U}^{^2} , {\tilde U}^{^3}]=\beta {\tilde U}^{^4},~~~~~~~~~{[U_{_1} , {\tilde U}^{^3}]}= -{\tilde U}^{^4},\nonumber\\
{[U_{_4} , {\tilde U}^{^2}]}&=& \beta U_{_3},~~~~~~~\{U_{_4} , {\tilde U}^{^3}\}=\beta U_{_2}-{\tilde U}^{^1},\label{5.11}
\end{eqnarray}
where $U_{_a}$'s generate the ${\C}^3 \oplus {\A}_{1,1}$,
while the bases ${\tilde U}^a$'s stand for the ${\C}^3 \oplus {\A}_{1,1}.i$. Also, $\beta$ is a non-zero constant.
Here and henceforth we denote the bosonic bases by $(U_{_1}, U_{_2}, {\tilde U}^{^1}, {\tilde U}^{^2})$, and fermionic ones by
$(U_{_3}, U_{_4}, {\tilde U}^{^3}, {\tilde U}^{^4})$.
The isomorphism between the superdoubles $\D=({\C}^3 \oplus {\A}_{1,1} , {\cal I}_{_{(2|2)}})$
and ${\D}_{_U}=({\C}^3 \oplus {\A}_{1,1} , {\C}^3 \oplus {\A}_{1,1}.i)$ is given by the following transformation:
\begin{eqnarray}
{T_{_1}} &=&U_{_1},~~~~~~~~~~~~~~{\tilde T}^{^1} =-\beta U_{_2} + {\tilde U}^{^1},\nonumber\\
{T_{_2}} &=&U_{_2},~~~~~~~~~~~~~~{\tilde T}^{^2} =\beta U_{_1} + {\tilde U}^{^2},\nonumber\\
{T_{_3}} &=& - U_{_3},~~~~~~~~~~~~{\tilde T}^{^3} =- {\tilde U}^{^3},\nonumber\\
{T_{_4}} &=& - U_{_4},~~~~~~~~~~~~{\tilde T}^{^4} =- {\tilde U}^{^4}.\label{5.12}
\end{eqnarray}
Comparing relations \eqref{5.2} and \eqref{5.12} one can find the sub-matrices ${F_{_a}}^{^c},  G_{_{ac}},  H^{^{bc}}$ and  ${K^{^b}}_{_{c}}$.
Then, using these and constant matrix $E_{0}$ of \eqref{3.70} together with formulae \eqref{5.5} and \eqref{5.6} we can find the matrices
$M_{_{ab}}$ and ${N^{^a}}_{_{b}}$ leading to
\begin{eqnarray}
NM^{^{-1}} = \left( \begin{array}{cccc}
e^{-y}  & -\beta  & 0 & 0\\
\beta  & -e^{-y}  & 0 & 0\\
0 & 0  & 0 & e^{-y}\\
0  & 0  & -e^{-y}& 0\\
\end{array} \right).\label{5.13}
\end{eqnarray}
We use the parametrization of a general group element of ${C}^3 \oplus {A}_{1,1}$ as in \eqref{3.54}, but with
$T_{_a}$'s replaced by $U_{_a}$'s. Then, one utilizes
formula \eqref{3.10} to calculate the matrices $a(g_{_U})$ and $b(g_{_U})$ for this decomposition, giving us
\begin{eqnarray}\label{5.14}
a_{_{a}}^{^{^{~b}}}(g_{_U}) =\left( \begin{array}{cccc}
1  & 0  & -\chi & 0\\
0  & 1  & 0 & 0\\
0 & 0  & -1 & 0\\
0  & 0  & {x_{_{1}}} & -1\\
\end{array} \right),~~~~~~~~~~b^{^{ab}}(g_{_U})=\left( \begin{array}{cccc}
0  & 0  & 0 & 0\\
0  & 0 & \beta \chi & 0\\
0 & -\beta \chi  & 0 & 0\\
0  & 0  & 0 & 0\\
\end{array} \right).
\end{eqnarray}
Plugging \eqref{5.14} into \eqref{5.7} one gets the Poisson superbracket ${\Pi}^{^{ab}}(g_{_U})$,
which the result is exactly similar to sub-matrix $b^{^{ab}}(g_{_U})$ of equation \eqref{5.14}.
These results give us the background
\begin{eqnarray}
{E}_{_{ab}}(g_{_U}) = \left( \begin{array}{cccc}
-\frac{e^{^{-y }}}{\Delta(y)} & \frac{\beta}{\Delta(y)}  & 0 & \frac{\beta^2 \chi e^{^{y }}}{\Delta(y)}\\
-\frac{\beta}{\Delta(y)}  & \frac{e^{^{-y }}}{\Delta(y)}  & 0 & \frac{\beta \chi }{\Delta(y)}\\
0 & 0  & 0 & e^{^{y }}\\
\frac{\beta^2 \chi e^{^{y }}}{\Delta(y)}  & -\frac{\beta \chi}{\Delta(y)}  & -e^{^{y}} & 0\\
\end{array} \right),\label{5.15}
\end{eqnarray}
where $\Delta(y) = \beta^2 -e^{-2y}$. In order to calculate the corresponding dilaton field we find that
$s\hspace{-0.6mm}\det(_a{{E}_{_{b}}(g_{_U})}) = e^{-2y}/\Delta(y),~s\hspace{-0.6mm}\det(_a{{M}_{_{b}}(g_{_U})}) =-1$ and
$s\hspace{-0.6mm}\det(a_{_{a}}^{^{^{~b}}}(g_{_U})) =1$. Then, the dilaton is obtained by making use of
\eqref{5.8}, and by remembering that $\phi^{^{(0)}}= {\cal C}_{_{0}}$ one gets the final result in the form
\begin{eqnarray}\label{5.16}
\Phi_{_U} = {\cal C}_{_{0}} -\frac{1}{2} \log\big|\beta^2 e^{2y}-1\big|.
\end{eqnarray}
Using \eqref{5.15} and the right-invariant super one-forms on the ${C}^3 \oplus {A}_{1,1}$ (equation
\eqref{3.69}) one can write the action of model on the superdouble $({\C}^3 \oplus {\A}_{1,1} , {\C}^3 \oplus {\A}_{1,1}.i)$
in the form of \eqref{5.3}.
Finally,  background including the line element and $B$-field is, in the coordinate basis, read off
\begin{eqnarray}
ds^2 &=& \frac{l^2}{4} dy^2 +\frac{1}{\Delta(y)} \Big[e^{-y} \big(-d{x_{_{1}}}^2 + d{x_{_{2}}}^2\big)
-2 {\beta}^2 e^y~ d{x_{_{1}}}  \chi d \chi\Big]-2 e^y ~d \psi  d \chi,  \nonumber\\
B&=&\frac{\beta}{\Delta(y)}\Big[d{x_{_{1}}} \wedge d{x_{_{2}}} -  \chi ~ d{x_{_{2}}} \wedge  d \chi\Big].\label{5.17}
\end{eqnarray}
By taking into consideration the scalar curvature of the metric which is
\begin{eqnarray}
{\cal R}=-\frac{8 \beta^2 (2 \beta^2 +5 e^{-2y})}{l^2 \Delta^{^2}(y)},\label{5.18}
\end{eqnarray}
one verifies the one-loop conformal invariance conditions of the background \eqref{5.17} with dilaton field \eqref{5.16} and a
zero cosmological constant. Note that the metric of \eqref{5.17} is ill defined at the region $y=-\ln|\beta|$.
This singularity also appears in the scalar curvature, and hence, $y=-\ln|\beta|$ is a true singularity.
In fact, we have shown that the solution of BTZ vacuum metric coupled to two fermionic fields
with no curvature singularity is, under the T-plurality, related to a solution with a curvature singularity.
This situation was not seen in the previous section regarding to super non-Abelian T-duality.
\\\\
\smallskip
$\bullet$~${\D}_{_U}={\bf \big({\C}^3 \oplus {\A}_{1,1} , {(2{\A}_{1,1}+2{ \A})^0}.i\big)}$:
From the Manin supertriples of \eqref{5.9} we see that $\D=({\C}^3 \oplus {\A}_{1,1} , {\cal I}_{_{(2|2)}})$ and
${\D}_{_U}={\big({\C}^3 \oplus {\A}_{1,1} , {(2{\A}_{1,1}+2{ \A})^0}.i\big)}$ as Lie
superalgebras are isomorphic, and so one can find an isomorphism transforming the Lie multiplication of $\D$ into that of ${\D}_{_{U}}$
and preserving the canonical form of the bilinear form $<. , .>$ such that they belong to the same Drinfeld superdouble.
The isomorphism transformation of Manin supertriples between $({\C}^3 \oplus {\A}_{1,1} , {\cal I}_{_{(2|2)}})$ and
${\big({\C}^3 \oplus {\A}_{1,1} , {(2{\A}_{1,1}+2{ \A})^0}.i\big)}$ is given by
\begin{eqnarray}
{T_{_1}} &=&U_{_1},~~~~~~~~~~~~~~{\tilde T}^{^1} = {\tilde U}^{^1},\nonumber\\
{T_{_2}} &=&U_{_2},~~~~~~~~~~~~~~{\tilde T}^{^2} ={\tilde U}^{^2},\nonumber\\
{T_{_3}} &=& - U_{_3},~~~~~~~~~~~~{\tilde T}^{^3} =-\frac{\gamma}{2} U_{_4}- {\tilde U}^{^3},\nonumber\\
{T_{_4}} &=& - U_{_4},~~~~~~~~~~~~{\tilde T}^{^4} =-\frac{\gamma}{2} U_{_3} - {\tilde U}^{^4}.\label{5.19}
\end{eqnarray}
Here, $\{U_{_a} , {\tilde U}^{^b}\}$ are elements of bases of the superdouble
${\big({\C}^3 \oplus {\A}_{1,1} , {(2{\A}_{1,1}+2{ \A})^0}.i\big)}$ whose Lie superalgebra  is defined by
the following non-zero (anti-)commutation relations:
\begin{eqnarray}
[U_{_1} , U_{_4}] = U_{_3},~~~~\{{\tilde U}^{^3} , {\tilde U}^{^3}\}=\gamma {\tilde U}^{^1},~~~~{[U_{_1} , {\tilde U}^{^3}]}=-\gamma U_{_3} -{\tilde U}^{^4},~~~~\{U_{_4} , {\tilde U}^{^3}\}=-{\tilde U}^{^1},~~\label{5.20}
\end{eqnarray}
where $\gamma$ is a non-zero constant.
In this case, the matrix $a_{_{a}}^{^{^{~b}}}(g_{_U})$ is the same form as in \eqref{5.14}.
Calculating the matrices $M_{_{ab}}$, ${N^{^a}}_{_{b}}$ and ${\Pi}^{^{ab}}(g_{_U})$ for the decomposition \eqref{5.19} we then get
\begin{eqnarray}
NM^{^{-1}} = \left( \begin{array}{cccc}
e^{-y}  & 0  & 0 & 0\\
0  & -e^{-y}  & 0 & 0\\
0 & 0  & 0 & (e^{-y} - \frac{\gamma}{2})\\
0  & 0  & -(e^{-y} + \frac{\gamma}{2})& 0\\
\end{array} \right),~~{\Pi}(g_{_U}) = \left( \begin{array}{cccc}
0  & 0  & 0 & 0\\
0  &0  & 0 & 0\\
0 & 0  & -\gamma x_{_{1}} & 0\\
0  & 0  & 0& 0\\
\end{array} \right),~~~\label{5.21}
\end{eqnarray}
leading to a background
\begin{eqnarray}\label{5.22}
{E}_{_{ab}}(g_{_U}) &=& \left( \begin{array}{cccc}
e^{y} & 0      & 0 & 0\\
0     & -e^{y} & 0 & 0\\
0     & 0      & 0 & \frac{1}{(e^{-y} + \frac{\gamma}{2})}\\
0     & 0      & -\frac{1}{(e^{-y} - \frac{\gamma}{2})} & \frac{\gamma x_{_{1}}}{(e^{-2y}- \frac{\gamma^2}{4})}\\
\end{array} \right).
\end{eqnarray}
In order to evaluate the total dilaton contribution we have to use $\phi^{^{(0)}}= {\cal C}_{_{0}}$
which gives the final result
\begin{eqnarray}
\Phi_{_U} = {\cal C}_{_{0}} + \frac{1}{2} \log\big|\frac{\gamma^2}{4} e^{2y}-1\big|.\label{5.23}
\end{eqnarray}
Since the first sub-superalgebra is ${\C}^3 \oplus {\A}_{1,1}$,
the right-invariant super one-forms are the same forms as in \eqref{3.69}.
Finally, using \eqref{5.3} one can construct the action of the model on the superdouble
$\big({\C}^3 \oplus {\A}_{1,1} , {(2{\A}_{1,1}+2{ \A})^0}.i\big)$
whose background is
\begin{eqnarray}
ds^2 &=& \frac{l^2}{4} dy^2 +e^{y}\big(d{x_{_{1}}}^2 - d{x_{_{2}}}^2\big)
-\frac{2e^{-y}}{(e^{-2y}- \frac{\gamma^2}{4})} ~d \psi  d \chi,  \nonumber\\
B&=&\frac{\gamma}{2(e^{-2y}- \frac{\gamma^2}{4})}~d\psi \wedge d\chi.\label{5.24}
\end{eqnarray}
One immediately finds that the scalar curvature of the metric is
\begin{eqnarray}
{\cal R}=\frac{\gamma^2 (10 e^{-2y} - \gamma^2)}{l^2 (e^{-2y} -\frac{\gamma^2}{4})^2}.\label{5.25}
\end{eqnarray}
Again we encounter a true singularity that appears here at the region $y=-\ln|{\gamma}/{2}|$.
Looking at the one-loop beta-function equations
one verifies the conformal invariance conditions of the background \eqref{5.24} with dilaton field \eqref{5.23} and a
zero cosmological constant.
\\
\\
\smallskip
$\bullet$~${\D}_{_U}={\bf \big({\C}^3 \oplus {\A}_{1,1} , \big(({\A}_{1,1}+2{\A})^2 \oplus {\A}_{1,1}\big).i\big)}$:
As the last example of this section we ask the question of super PL T-plurality of the BTZ vacuum metric coupled to two fermionic fields
with respect to the superdouble $\big({\C}^3 \oplus {\A}_{1,1} , \big(({\A}_{1,1}+2{\A})^2 \oplus {\A}_{1,1}\big).i\big)$ whose Lie superalgebra
is defined by the following non-zero Lie superbrackets:
\begin{eqnarray}
[U_{_1} , U_{_4}] &=&U_{_3},~~~~~~~~~~~\{{\tilde U}^{^3} , {\tilde U}^{^4}\}=\gamma {\tilde U}^{^1},~~~~~~~~~{[U_{_1} , {\tilde U}^{^3}]}= - \gamma U_{_4}
-{\tilde U}^{^4},\nonumber\\
{[U_{_1} , {\tilde U}^{^4}]}&=& -\gamma U_{_3},~~~~~~~\{U_{_4} , {\tilde U}^{^3}\}=-{\tilde U}^{^1},\label{5.26}
\end{eqnarray}
where $\gamma$ is a non-zero constant. The isomorphism transforming the Lie multiplication
of $\D=({\C}^3 \oplus {\A}_{1,1} , {\cal I}_{_{(2|2)}})$ into that of
${\D}_{_{U}} = \big({\C}^3 \oplus {\A}_{1,1} , \big(({\A}_{1,1}+2{\A})^2 \oplus {\A}_{1,1}\big).i\big)$
and preserving the canonical form of the bilinear form $<. , .>$,
is given by
\begin{eqnarray}
{T_{_1}} &=&U_{_1},~~~~~~~~~~~~~~{\tilde T}^{^1} = {\tilde U}^{^1},\nonumber\\
{T_{_2}} &=&U_{_2},~~~~~~~~~~~~~~{\tilde T}^{^2} ={\tilde U}^{^2},\nonumber\\
{T_{_3}} &=& - U_{_3},~~~~~~~~~~~~{\tilde T}^{^3} =- {\tilde U}^{^3},\nonumber\\
{T_{_4}} &=& - U_{_4},~~~~~~~~~~~~{\tilde T}^{^4} =-{\gamma} U_{_4} - {\tilde U}^{^4}.\label{5.27}
\end{eqnarray}
After calculating the sub-matrices ${F_{_a}}^{^c},  G_{_{ac}},  H^{^{bc}}$ and  ${K^{^b}}_{_{c}}$ we employ formulae \eqref{5.5} and \eqref{5.6}
to obtain the matrices $M_{_{ab}}$ and ${N^{^a}}_{_{b}}$, in such a way that one must also use constant matrix $E_{0}$ of \eqref{3.70}.
In addition, the matrix ${\Pi}(g_{_U})$ is obtained for this decomposition by using \eqref{3.10} and \eqref{5.7}.
It then follows from \eqref{5.4} that
\begin{eqnarray}\label{5.28}
{E}_{_{ab}}(g_{_U}) &=& \left( \begin{array}{cccc}
 e^y & 0  & 0 &0\\
0 & -e^y  & 0 & 0\\
0  & 0  & \gamma e^{2y} & -(\gamma x_{_{1}} e^{2y}- e^y)\\
0  & 0  & -(\gamma x_{_{1}} e^{2y}+ e^y) & \gamma {x_{_{1}}}^2 e^{2y}\\
\end{array} \right).
\end{eqnarray}
Finally, background including the line element and $B$-field takes the following form
\begin{eqnarray}
ds^2&=& \frac{l^2}{4}  dy^2 + e^y \Big[{d x_{_{1}}}^2  - ~ {d x_{_{2}}}^2 -2  ~ d\psi d\chi\Big],\nonumber\\
B&=& -\frac{1}{2} \gamma ~ e^{2y}~d\psi \wedge d\psi.\label{5.29}
\end{eqnarray}
Unlike the previous two examples, in this case the metric is flat in the sense that both its scalar curvature and Ricci tensor vanish.
In addition, it is interesting to note that the above metric is exactly the same as the metric of action \eqref{5.10}
on the superdouble $({\C}^3 \oplus {\A}_{1,1} , {\cal I}_{_{(2|2)}})$.
Here, the dilaton obtained from equation \eqref{5.8} is ${\Phi} = \phi^{^{(0)}}$.
Finally we get the dilaton by remembering that $\phi^{^{(0)}}= {\cal C}_{_0}$ which gives the final result ${\Phi}={\cal C}_{_0}$.

\section{Conclusion}
\label{Sec.VI}
In summary, we have studied Abelian T-duality, non-Abelian T-duality and T-plurality of the BTZ metric coupled to two fermionic fields.
In studying Abelian T-duality, we have obtained two duals for the BTZ metric when is coupled to two fermionic fields.
When the dualizing was implemented by the Abelian Lie supergroup of the type $(2|2)$, we found that the singularity of the dual metric
is appeared at the region $r = \frac{J}{2} M^{\frac{-1}{2}}$,
whereas when we dealt with a $(2|1)$-dimensional Abelian Lie supergroup, we encountered a singularity at the origin.
Moreover, in both cases of the dual metrics, horizons were at the same location as the metric of original model.

Using super PL T-duality approach in the presence of spectator fields we have constructed some
non-Abelian T-dualizable $\sigma$-models on some of the semi-Abelian Drinfeld superdoubles,
in such a way that we have dealt with $(2|2)$-dimensional non-Abelian Lie superalgebras
$({\C}_1^1 +{\A})$, $({\C}^3 +{\A})$, ${\C}^3 \oplus {\A}_{1,1}$,
${\D}^{10}_{p=\pm{1/2}}$,  and $(2{ \A}_{1,1}+2{ \A})^0$ as the first sub-superalgebras of the superdoubles.
By a convenient choice of the spectator-dependent background matrices we showed that the background of original $\sigma$-models describes
a string propagating in a target space with the BTZ vacuum metric coupling to two fermionic fields,
in such a way that a new family of the solutions to supergravity equations was found in both cases of absence
and presence of the field strength.
Accordingly, we think that the choice of spectator-dependent matrices plays a key role in the structure of our models.
In some cases we showed that the dual backgrounds also yields the BTZ vacuum metric coupled to two fermionic fields.

Starting from the decomposition of semi-Abelian Drinfeld superdouble $({\C}^3 \oplus {\A}_{1,1} , {\cal I}_{_{(2|2)}})$
we could calculate the super PL T-plural of the BTZ vacuum metric coupled to two fermionic fields with respect to the ${C}^3 \oplus {A}_{1,1}$  Lie supergroup.
In fact, we obtained a conformally invariant duality chain of $2+1$-dimensional
cosmological string backgrounds coupled to two fermionic fields in the form
of relations \eqref{5.10}, \eqref{5.17}, \eqref{5.24} and \eqref{5.29}.
The most interesting feature of our super PL T-plural $\sigma$-models is the existence of true singularities in some of the metrics of the models.
This situation was not seen in the case of super non-Abelian T-duals.
Our current goal was to get better understanding of super PL
T-plurality through investigation of examples presented in section \ref{Sec.V}.
We hope that the discussed super PL T-plurality can yield insights to supergravity.

The findings of our study showed that the target supermanifold $\M \approx O \times G$ with non-Abelian Lie supergroups of
the type $(2|2)$ is wealthy.
In addition to super PL symmetric backgrounds constructed out in this
paper one can study the super non-Abelian T-dualization of Schwarzschild and G\"{o}del metrics when are coupled
to two fermionic fields.
We intend to address this problem in the future.
\appendix

\section{Lie superalgebras of the type $(2 | 2)$}

The indecomposable Lie superalgebras of the type $(2 | 2)$ was first presented by Backhouse in \cite{B}.
In that classification, Lie superalgebras are divided
into two types: trivial and nontrivial Lie superalgebras for which the anti-commutations of F-F are,
respectively, zero or non-zero.
The decomposable Lie superalgebras of the type $(2 | 2)$ have been recently
obtained in \cite{ER6}.
As mentioned in Introduction section, there are some omissions and redundancies in the Backhouse's classification, probably because
the author in some cases has not taken into account all isomorphisms to reduce the list of superalgebras.
In this Appendix we make some modifications to the Backhouse's classification.
Before proceeding to this end, it should be noted that Lie superalgebras of the type $(2 | 2)$ possess two bosonic
generators $\{T_{_1}, T_{_2}\}$ and two fermionic ones $\{T_{_3}, T_{_4}\}$ as was mentioned in
section \ref{Sec.IV}.

Below one can find the Lie superalgebras of Backhouse's classification that are isomorphic together
\begin{eqnarray}\label{A.1}
&& (2{\A}_{1,1}+2{ \A})^1 \cong  (2{\A}_{1,1}+2{ \A})^2 \cong  {(2{\A}_{1,1}+2{ \A})}_{0<p<1/2}^3 \cong  {(2{\A}_{1,1}+2{ \A})}_{p>0}^4,~~~~ \label{A.1}\\
&& {(2{\A}_{1,1}+2{ \A})}_{p>1/2}^3 \cong  {(2{\A}_{1,1}+2{ \A})}_{p=1}^3. \label{A.2}
\end{eqnarray}
However, one can check that the following real transformations guarantee the isomorphism between these Lie superalgebras:\\
$\bullet$~One can simply show that under the transformation
\begin{eqnarray}\label{A.3}
	T'_{_1} & = & {a^2} T_{_1},~~~~~~~~~~~~~~~~~~~~ T'_{_3} = -a T_{_3}, \nonumber\\
	T'_{_2} & = & {a^2} T_{_1} +b^{^2}  T_{_2},~~~~~~~~~~~T'_{_4} = -a T_{_3} -b T_{_4},
\end{eqnarray}
for some constants $a, b$, the Lie superalgebras $(2{\A}_{1,1}+2{ \A})^1$ and $(2{\A}_{1,1}+2{ \A})^2$ are isomorphic together, such that
$T_{_a}$'s,  $a=1,...,4$ are the bases of the $(2{\A}_{1,1}+2{ \A})^1$, while $T'_{_a}$'s,  $a=1,...,4$ stand for the $(2{\A}_{1,1}+2{ \A})^2$.\\
$\bullet\bullet$~The isomorphism of between the Lie superalgebras $(2{\A}_{1,1}+2{ \A})^1$ and ${(2{\A}_{1,1}+2{ \A})}_{p>0}^4$ is given by the following transformation
\begin{eqnarray}\label{A.4}
	T'_{_1} & = & \lambda_{_\pm}^2 ~T_{_1}+T_{_2},~~~~~~~~~~~~~~~~~~ T'_{_3} = \lambda_{_\pm} T_{_3} -  T_{_4}, \nonumber\\
	T'_{_2} & = &  T_{_1} +\lambda_{_\pm}^2 ~ T_{_2},~~~~~~~~~~~~~~~~~~T'_{_4} = -T_{_3} - \lambda_{_\pm} T_{_4},
\end{eqnarray}
where $\lambda_{_\pm} =\frac{1}{2p} (-1\pm \sqrt{1+4p^2})$.
Here, $T_{_a}$'s, $a=1,...,4$ and $T'_{_a}$'s, $a=1,...,4$ are generators of the $(2{\A}_{1,1}+2{ \A})^1$ and
${(2{\A}_{1,1}+2{ \A})}_{p>0}^4$, respectively.\\
$\bullet\bullet\bullet$~
One may use the transformation
\begin{eqnarray}\label{A.5}
	T'_{_1} & = & (\frac{\delta}{p} -2p \delta -1) T_{_1}+ (1-2p \delta) T_{_2},~~~~~~~~~~~~~~~~ T'_{_3} = \delta T_{_3} -  T_{_4}, \nonumber\\
	T'_{_2} & = &  (1-2p \delta) T_{_1} +(\frac{\delta}{p} -2p \delta -1) T_{_2},~~~~~~~~~~~~~~~~T'_{_4} = -T_{_3} + \delta T_{_4},
\end{eqnarray}
to show that the Lie superalgebras ${(2{\A}_{1,1}+2{ \A})}_{0<p<1/2}^3$ and $(2{\A}_{1,1}+2{ \A})^1$
are isomorphic together. Here, we have used the bases $\{T_{_a}\}$  and $\{T'_{_a}\}$ to define the ${(2{\A}_{1,1}+2{ \A})}_{0<p<1/2}^3$ and $(2{\A}_{1,1}+2{ \A})^1$, respectively. Note that the parameter $\delta$ in \eqref{A.5} is given by $\delta=\frac{1}{2p} (1+\sqrt{1-4p^2})$.\\
$\bullet\bullet\bullet\hspace{1mm}\bullet$~ The isomorphism transforming the Lie superalgebra of ${(2{\A}_{1,1}+2{ \A})}_{p=1}^3$ into that of
${(2{\A}_{1,1}+2{ \A})}_{p>1/2}^3 $ is expressed as
\begin{eqnarray}\label{A.6}
T'_{_1} & = & \Big[-1 + 3p^2 + (1-2p^2) \gamma_{_\pm} \Big] T_{_1}+ p^2 (1-2\gamma_{_\pm}) T_{_2},~~~ T'_{_3} = -\gamma_{_\pm} T_{_3} +p T_{_4}, \nonumber\\
T'_{_2} & = &  (1-4p^2) T_{_1},~~~~~~~~~~~~~~~~~~~~~~~~~~~~~~~~~~~~~~~~~~~~~~~~~T'_{_4} = -T_{_3} + 2p T_{_4},
\end{eqnarray}
where $\gamma_{_\pm} =\frac{1}{2} \big[1 \pm \sqrt{3(4p^2-1)}\big]$.
Here, the Lie superalgebras  ${(2{\A}_{1,1}+2{ \A})}_{p>1/2}^3 $ and ${(2{\A}_{1,1}+2{ \A})}_{p=1}^3$ are generated by means of the bases
$\{T_{_a}\}$  and $\{T'_{_a}\}$, respectively.

We should also pay attention to the fact that Lie superalgebra $({\D}^7_{1-p\;p})$ for the value of $p=0$ is isomorphic to that of
$({\C}^1_1+{\A})$; moreover, it is isomorphic to the ${({\D}^7_{\frac{1}{2}\;\frac{1}{2}})}^2$ if one sets $p=1/2$.
We have included the above modifications to the Backhouse's classification, and
have classified all decomposable and indecomposable Lie superalgebras of the type $(2|2)$ into six disjoint families in Tables 2 to 7.
Our classification is based on the (anti-)commutation relations between bases of B-B, B-F and F-F.
Note that in Tables 2 to 7, only Lie superalgebras having non-zero (anti-)commutation relations have been listed.
Accordingly, the Abelian Lie superalgebra ${\cal {I}}_{_{(2|2)}}$ is absent in these Tables.

\newpage
\begin{center}
		\small{{{\bf Table 2.}~ Familiy I: (B-B, B-F, F-F)}}
		{\scriptsize
			\renewcommand{\arraystretch}{1.5}{
\begin{tabular}{| l|l|l|} \hline \hline
{\scriptsize $ \G$ }& {\scriptsize Non-zero (anti-)commutation relations} & {\scriptsize Comments} \\ \hline
 {\scriptsize ${({\D}^7_{\frac{1}{2}\;\frac{1}{2}})}^1$}& {\scriptsize
$[T_1,T_2]=T_2, \;\;[T_1,T_3]=\frac{1}{2}
T_3,\;\;[T_1,T_4]=\frac{1}{2}
T_4,\;\; \{T_3,T_3\}=T_2,$}  & \\
& {\scriptsize $\{T_4,T_4\}=T_2$} & \\

{\scriptsize ${({\D}^7_{\frac{1}{2}\;\frac{1}{2}})}^2$}& {\scriptsize
$[T_1,T_2]=T_2, \;\;[T_1,T_3]=\frac{1}{2}
T_3,\;\;[T_1,T_4]=\frac{1}{2}
T_4,\;\; \{T_3,T_3\}=T_2,$} & \\
& {\scriptsize
$ \{T_4,T_4\}=-T_2$} &\\

{\scriptsize ${({\D}^7_{\frac{1}{2}\;\frac{1}{2}})}^3$}& {\scriptsize
$[T_1,T_2]=T_2, \;\;[T_1,T_3]=\frac{1}{2}
T_3,\;\;[T_1,T_4]=\frac{1}{2}
T_4,\;\; \{T_3,T_3\}=T_2 $} &\\

{\scriptsize $({\D}^7_{1-p\;p})$ }& {\scriptsize $[T_1,T_2]=T_2,
\;\;[T_1,T_3]=p T_3,\;\;[T_1,T_4]=(1-p)
T_4,\;\{T_3,T_4\}=T_2 $}& {\scriptsize $p\neq0,~p < \frac{1}{2}$} \\

{\scriptsize $({\D}^8_{\frac{1}{2}})$ }& {\scriptsize $[T_1,T_2]=T_2,
\;\;[T_1,T_3]=\frac{1}{2} T_3,\;\;[T_1,T_4]=T_3+\frac{1}{2}
T_4,\; \{T_4,T_4\}=T_2 $} &\\

{\scriptsize $({\D}^9_{\frac{1}{2}\;p})$ }& {\scriptsize
$[T_1,T_2]=T_2, \;\;[T_1,T_3]=\frac{1}{2} T_3-p
T_4,\;\;[T_1,T_4]=p T_3+\frac{1}{2} T_4,$}&
 {\scriptsize $p > 0$}  \\
& {\scriptsize $\{T_3,T_3\}=T_2,\;\; \{T_4,T_4\}=T_2$}& \\

{\scriptsize $({\D}^{10}_{0})^1$ }& {\scriptsize $[T_1,T_2]=T_2,
\;\;[T_1,T_3]= T_3, \;\;[T_2,T_4]= T_3, \;\; \{T_4,T_4\}=T_1,$} &\\
&{\scriptsize $\{T_3,T_4\}=-\frac{1}{2}T_2$}&\\

{\scriptsize $({\D}^{10}_{0})^2$ }& {\scriptsize $[T_1,T_2]=T_2,
\;\;[T_1,T_3]= T_3, \;\;[T_2,T_4]= T_3, \;\; \{T_4,T_4\}=-T_1,$} &\\
&{\scriptsize $\{T_3,T_4\}=\frac{1}{2}T_2$}&\\

{\scriptsize $({\C}^1_1+{\A})$ }& {\scriptsize $[T_1,T_2]=T_2,
\;\;[T_1,T_3]=T_3,
\;\;\{T_3,T_4\}=T_2 $} &\\

{\footnotesize $({\C}^1_{\frac{1}{2}} + {\A})$ }& {\footnotesize
	$[T_1,T_2]=T_2,\; \;[T_1,T_3]=\frac{1}{2}T_3,\; \;\{T_3,T_3\}=T_2
	$}&{\scriptsize $\equiv({\C}^1_{\frac{1}{2}}) \oplus {\A}$ } \\
\hline\hline
		\end{tabular}}}
\end{center}

\vspace{4mm}
\begin{center}
\small{{{\bf Table 3.}~ Familiy II: (B-F, F-F)}}
		{\scriptsize
			\renewcommand{\arraystretch}{1.5}{
\begin{tabular}{| l|l|l|} \hline \hline
{\scriptsize $ \G$ }& {\scriptsize Non-zero (anti-)commutation relations} & {\scriptsize Comments} \\ \hline

{\scriptsize $({\C}^2_{-1}+{\A})$} & {\scriptsize $[T_1,T_3]=T_3,
	\;\;[T_1,T_4]=-T_4,
	\;\;\{T_3,T_4\}=T_2 $}& {\scriptsize Jordan-Winger }\\
& & {\scriptsize quantization }\\

{\scriptsize $({\C}^3+{\A})$ }& {\scriptsize $[T_1,T_4]=T_3,
	\;\;\{T_4,T_4\}=T_2 $}&{\scriptsize Nilpotent}\\

{\scriptsize $({\C}^5_0+{\A})$ }& {\scriptsize $[T_1,T_3]=-T_4,
	\;[T_1,T_4]=T_3,
	\;\{T_3,T_3\}=T_2,\;\{T_4,T_4\}=T_2 $} &\\

{\scriptsize $\big({\B}+({\A}_{1,1}+{\A})\big)$ }& {\scriptsize $[T_1,T_3]=T_3,\;\;\{T_4,T_4\}=T_2 $} &\\
\hline\hline
\end{tabular}}}
\end{center}
\vspace{4mm}

\begin{center}
\small{{{\bf Table 4.}~ Familiy III: (B-F)}}
		{\scriptsize
			\renewcommand{\arraystretch}{1.5}{
\begin{tabular}{| l|l|l|} \hline \hline
{\scriptsize $ \G$ }& {\scriptsize Non-zero (anti-)commutation relations} & {\scriptsize Comments} \\ \hline
	
{\scriptsize ${\D}^5~~$}& {\scriptsize $[T_1,T_3]=T_3, \;\;[T_1,T_4]=T_4,\;\;[T_2,T_4]=T_3$}& \\

{\scriptsize ${\D}^6$}&{\scriptsize
	$[T_1,T_3]=T_3, \;\;[T_1,T_4]=T_4,\;\;[T_2,T_3]=-T_4,\;\;[T_2,T_4]=T_3 $} &\\

{\footnotesize ${\B} \oplus {\B}$}& {\footnotesize $[T_1,T_3]=T_3,
	\;\;[T_2,T_4]=T_4$} &\\

{\footnotesize ${\B} \oplus {\A} \oplus {\A}_{1,1}$}& {\footnotesize
	$[T_1,T_3]=T_3$}&{\scriptsize
	$\equiv {\C}^2_{p=0} \oplus {\A}_{1,1}$} \\

{\footnotesize ${\C}^2_p \oplus {\A}_{1,1}$}& {\footnotesize
	$[T_1,T_3]=T_3,~~[T_1,T_4]=pT_4$}& {\scriptsize$0< |p| \leq 1 $} \\

{\footnotesize ${\C}^3 \oplus {\A}_{1,1}$}& {\footnotesize
	$[T_1,T_4]=T_3$}& {\scriptsize Nilpotent} \\

{\footnotesize ${\C}^4 \oplus {\A}_{1,1}$}& {\footnotesize
	$[T_1,T_3]=T_3,\;\; [T_1,T_4]=T_3+T_4$}& \\

{\footnotesize ${\C}^5_p \oplus {\A}_{1,1}$}& {\scriptsize
	$[T_1,T_3]=pT_3-T_4,~~\;~~~[T_1,T_4]=T_3+pT_4$}&{\scriptsize
	$p \geq 0 $} \\
\hline\hline
\end{tabular}}}
\end{center}

\vspace{5mm}
	
\begin{center}
\small{{{\bf Table 5.}~ Familiy IV: (B-B, B-F)}}
		{\scriptsize
			\renewcommand{\arraystretch}{1.5}{
\begin{tabular}{| l|l|l|} \hline \hline
{\scriptsize $ \G$ }& {\scriptsize Non-zero (anti-)commutation relations} & {\scriptsize Comments} \\ \hline
	
	{\scriptsize ${\D}^1_{pq}$}&{\scriptsize
		$[T_1,T_2]=T_2, \;\;[T_1,T_3]=pT_3,\;\;[T_1,T_4]=qT_4 $}& {\scriptsize $pq \neq 0,\;\; p\geq q$}\\
	
	{\scriptsize ${\D}^8_{p}$}&{\scriptsize
		$[T_1,T_2]=T_2, \;\;[T_1,T_3]=pT_3,\;\;[T_1,T_4]=T_3+pT_4 $}& {\scriptsize $p \neq 0$}\\
	
	{\scriptsize ${\D}^9_{pq}$}&{\scriptsize
		$[T_1,T_2]=T_2, \;\;[T_1,T_3]=pT_3-qT_4,\;\;[T_1,T_4]=qT_3+pT_4 $}& {\scriptsize $q > 0$}\\
	
	{\scriptsize ${\D}^{10}_{p}$}&{\scriptsize
		$[T_1,T_2]=T_2, \;\;[T_1,T_3]=(p+1)T_3,\;\;[T_1,T_4]=pT_4,\;\;[T_2,T_4]=T_3 $} &\\
	
{\footnotesize ${\C}^1_{p} \oplus {\A}$}& {\footnotesize
	$[T_1,T_2]=T_2, \;\;[T_1,T_3]=p T_3$} &{\footnotesize $p \neq 0$}\\
\hline\hline
\end{tabular}}}
\end{center}
\vspace{4mm}


\begin{center}
\small{{{\bf Table 6.}~ Familiy V: (F-F)}}
		{\scriptsize
			\renewcommand{\arraystretch}{1.5}{
\begin{tabular}{| l|l|l|} \hline \hline
{\scriptsize $ \G$ }& {\scriptsize Non-zero (anti-)commutation relations} & {\scriptsize Comments} \\ \hline
	
{\scriptsize ${(2{\A}_{1,1}+2{\A})}^3_{p=1}$ }& {\scriptsize
	$\{T_3,T_3\}=T_1,~ \{T_4,T_4\}=T_2,~\{T_3,T_4\}=T_1+T_2  $} & {\scriptsize  Nilpotent}\\
	
{\scriptsize ${(2{\A}_{1,1}+2{\A})}^3_{p=\frac{1}{2}}$ }& {\scriptsize $\{T_3,T_3\}=T_1,~\{T_4,T_4\}=T_2,~\{T_3,T_4\}=\frac{1}{2}(T_1+T_2)$}
& {\scriptsize  Nilpotent}\\

{\footnotesize $(2{\A}_{1,1}+2{\A})^0$ }& {\footnotesize $\{T_3,T_3\}=T_1 $}&
{\scriptsize $\equiv({\A}_{1,1}+{\A})$}\\
& & {\scriptsize $\oplus {\A} \oplus {\A}_{1,1}$, Nilpotent}\\
\vspace{-1mm}

{\footnotesize $(2{\A}_{1,1}+2{\A})^1$ }& {\footnotesize $\{T_3,T_3\}=T_1,\; \;\; \{T_4,T_4\}=T_2  $}&
{\scriptsize $\equiv({\A}_{1,1}+{\A})$}\\
\vspace{-0.5mm}
& & {\scriptsize $\oplus ({\A}_{1,1}+{\A})$, Nilpotent}\\

{\scriptsize $\big(({\A}_{1,1}+2{\A})^1 \oplus {\A}_{1,1}\big) $ }& {\footnotesize
	$\{T_3,T_3\}=T_1,\; \;\; \{T_4,T_4\}=T_1  $}& {\scriptsize Nilpotent}\\

{\scriptsize $\big(({\A}_{1,1}+2{\A})^2 \oplus {\A}_{1,1}\big) $ }& {\footnotesize
	$\{T_3,T_3\}=T_1,\; \;\; \{T_4,T_4\}=-T_1  $}& {\scriptsize Nilpotent}\\
\hline\hline
\end{tabular}}}
\end{center}

\vspace{2mm}
	
\begin{center}
\small{{{\bf Table 7.}~ Familiy VI: (B-B)}}\\
		{\scriptsize
			\renewcommand{\arraystretch}{1.5}{
\begin{tabular}{| l|l|l|} \hline \hline
{\scriptsize $ \G$ }& {\scriptsize Non-zero (anti-)commutation relations~~~~} & {\scriptsize Comments} \\ \hline
	
{\footnotesize ${\C}^1_{p=0} \oplus {\A}~~~~~$} &{\footnotesize $[T_1,T_2]=T_2~~~~~~~~$}  & \\
\hline\hline
\end{tabular}}}
\end{center}

\vspace{2mm}

\section{Solutions of the super Jacobi and mixed super Jacobi
identities for the ${\C}^3 \oplus {\A}_{1,1}$ Lie superalgebra }

In order to obtain the Lie superbialgebra structures on the $\G={\C}^3 \oplus {\A}_{1,1}$ Lie superalgebra one must solve
the mixed super Jacobi identities \eqref{3.3} and also the following super Jacobi identity (on the dual Lie superalgebra ${\tilde \G}$)
\begin{equation}
(-1)^{a(b+c)}{\tilde{f}^{bd}}_{\; \; \; \; e}{\tilde{f}^{ca}}_{\; \; \; \; d}+
{\tilde{f}^{ad}}_{\; \; \; \; e}{\tilde{f}^{bc}}_{\; \; \; \; d}+(-1)^{c(a+b)}{\tilde{f}^{cd}}_{\; \; \; \; e}{\tilde{f}^{ab}}_{\; \; \; \; d}=0.
\label{B.1}
\end{equation}
Because of tensorial form of the
identities \eqref{3.3} and \eqref{B.1}, it is not easy to work with them.
So, we suggest writing these equations as matrix forms using the adjoint representations as presented in \cite{ER1}.
The initial steps of the analysis for solving the equations are made using a computer.
Thus, by solving equations \eqref{3.3} and \eqref{B.1} we obtain the general form of the structure constants related to the  dual  Lie superalgebras to
${\C}^3 \oplus {\A}_{1,1}$. The possibilities found by means of the computer are given by the following five cases:\\

\begin{tabular}{l l l l  p{15mm} }
\vspace{2mm}
{\it $Case$ $(1):$ }& {  ${{\tilde{f}}^{23}}_{\; \; \: 3}=\alpha,\;\;\;\;~~~~
 {{\tilde{f}}^{23}}_{\; \; \: 4} = \beta,\;\;~~\;\;{{\tilde{f}}^{24}}_{\; \; \: 4}= \gamma,
 \;\;\;\;~~{{\tilde{f}}^{12}}_{\; \; \: 1}=\gamma-\alpha.$}   \smallskip\\

\vspace{2mm}

{\it $Case$ $(2):$ }& {  ${{\tilde{f}}^{23}}_{\; \; \: 3}=-\alpha,\;\;\;\;~~
 {{\tilde{f}}^{23}}_{\; \; \: 4} = \beta,\;\;~~\;\;{{\tilde{f}}^{24}}_{\; \; \: 4}= \alpha,
 \;\;\;\;~~{{\tilde{f}}^{12}}_{\; \; \: 1}=2 \alpha,~~~~~~{{\tilde{f}}^{33}}_{\; \; \: 1}=\gamma.$}   \smallskip\\

\vspace{2mm}

{\it $Case$ $(3):$ }& {  ${{\tilde{f}}^{23}}_{\; \; \: 4}=\alpha,\;\;\;\;~~~~
 {{\tilde{f}}^{33}}_{\; \; \: 1} = \beta,\;\;~~\;\;{{\tilde{f}}^{34}}_{\; \; \: 1}= \gamma,
 \;\;\;\;~~{{\tilde{f}}^{12}}_{\; \; \: 1}=\frac{2 \alpha \gamma}{\beta},~~~~~~
 {{\tilde{f}}^{23}}_{\; \; \: 3}=-\frac{2 \alpha \gamma}{\beta}.$}   \smallskip\\
\vspace{2mm}

{\it $Case$ $(4):$ }& {  ${{\tilde{f}}^{33}}_{\; \; \: 1}=\alpha,\;\;\;\;~~~~
 {{\tilde{f}}^{33}}_{\; \; \: 2} = \beta,\;\;~~\;\;{{\tilde{f}}^{34}}_{\; \; \: 1}= \gamma,
 \;\;\;\;~~{{\tilde{f}}^{44}}_{\; \; \: 1}=\lambda.$}   \smallskip\\

 \vspace{2mm}

{\it $Case$ $(5):$ }& {  ${{\tilde{f}}^{24}}_{\; \; \: 4}=\alpha,\;\;\;\;~~~~
{{\tilde{f}}^{34}}_{\; \; \: 1} = \beta,\;\;~~\;\;{{\tilde{f}}^{44}}_{\; \; \: 1}= \gamma,
\;\;\;\;~~{{\tilde{f}}^{12}}_{\; \; \: 1}=-2 \alpha,  \;\;\;\;~~{{\tilde{f}}^{23}}_{\; \; \: 3}=3 \alpha,$}\\

 & {  ${{\tilde{f}}^{23}}_{\; \; \: 4}=- \frac{2 \alpha \beta}{\gamma},\;\;\;\;~~
{{\tilde{f}}^{33}}_{\; \; \: 1}= \frac{\beta^2}{\gamma}.$}   \smallskip\\
\end{tabular}\\
Using the formula of isomorphism transformation presented in \cite{ER1} (see also \cite{{ER6},{ER14}}) one can simply show that
the above dual solutions are isomorphic with some of the Lie superalgebras listed in Tables 2 to 7.
Finally, one may employ the automorphism Lie supergroup of ${\C}^3 \oplus {\A}_{1,1}$ and use the method demonstrated in \cite{ER1}
to obtain all inequivalent Lie superbialgebra structures on ${\C}^3 \oplus {\A}_{1,1}$.

\acknowledgments

This work has been supported by the research vice chancellor of Azarbaijan Shahid Madani University under research fund No. 1402/537.


\end{document}